\documentclass[aps,pre, preprint, showpacs,superscriptaddress]{revtex4-1}

\usepackage[english]{babel}
\usepackage{amsmath}
\usepackage{graphicx}
\usepackage{setspace}
\usepackage{bm}
\usepackage{color}
\usepackage{amssymb}
\usepackage{dsfont}
\usepackage{hyperref}
\usepackage[normalem]{ulem}

\usepackage{color}
\definecolor{azulUC3M}{RGB}{0,0,102}
\definecolor{gray97}{gray}{.97}
\definecolor{gray75}{gray}{.75}
\definecolor{gray45}{gray}{.45}
\definecolor{antiquebrass}{rgb}{0.8, 0.58, 0.46}
\definecolor{brickred}{rgb}{0.8, 0.25, 0.33}

\definecolor{hreflinkcolor}{rgb}{0.13,0.17,0.83}
\hypersetup{colorlinks=true,urlcolor=hreflinkcolor,
        linkcolor=hreflinkcolor,citecolor=hreflinkcolor}

\newcommand{\reffig} [1] {Fig.~\ref{#1}}

\definecolor{green2}{rgb}{0.0, 0.5, 0.0}

\newcommand{\EBedit}[1]{{\color{black} #1}}
\newcommand{\ESedit}[1]{{\color{black} #1}}
\newcommand{\GSAedit}[1]{{\color{black} #1}}
\newcommand{\TPedit}[1]{{\color{black} #1}}

\allowdisplaybreaks

\begin{document}
\singlespace
\title{Structure and evolution of magnetohydrodynamic solitary waves with Hall and finite Larmor radius effects}

\author{E.~Bello-Ben\'itez }
\email{ebello@ing.uc3m.es } \affiliation{Bioengineering
and Aerospace Engineering Department, Universidad Carlos III de
Madrid, Legan\'es, Spain}

\author{G.~S\'anchez-Arriaga}
\email{gonzalo.sanchez@uc3m.es} \affiliation{Bioengineering and
Aerospace Engineering Department, Universidad Carlos III de Madrid,
Legan\'es, Spain}

\author{T.~Passot}
\email{thierry.passot@oca.eu } \affiliation{Universit\'e C\^ote d'Azur, Observatoire de la  C\^ote d'Azur, CNRS, Laboratoire Lagrange, Bd de l'Observatoire, CS 34229, 06304 Nice cedex 4, France}

\author{D.~Laveder}\email{dimitri.laveder@oca.eu } \affiliation{Universit\'e C\^ote d'Azur, Observatoire de la  C\^ote d'Azur, CNRS, Laboratoire Lagrange, Bd de l'Observatoire, CS 34229, 06304 Nice cedex 4, France}

\author{E.~Siminos}\email{evangelos.siminos@physics.gu.se} \affiliation{Department of Physics, University of Gothenburg, 412 96 Gothenburg, Sweden}

\begin{abstract}

Nonlinear and low-frequency solitary waves are investigated in the framework of the one-dimensional  Hall-magnetohydrodynamic model with finite Larmor effects and a double adiabatic model for  plasma pressures. The organization of these localized structures in terms of the propagation angle with respect to the ambient magnetic field $\theta$ and the propagation velocity $C$ is discussed. There are three types of regions in the $\theta-C$ plane that correspond to domains where \ESedit{either} solitary waves cannot exist, are organized in branches, \ESedit{or} have a continuous spectrum. A numerical method valid for the two latter cases, that rigorously proves the existence of the waves, is presented and used to locate many waves, including bright and dark structures. Some of them belong to parametric domains where solitary waves were not found in previous works. The stability of the structures has been investigated by first performing a linear analysis of the background plasma state and second by means of numerical simulations. They show that the cores of some waves can be robust but, for the parameters considered in the analysis, the tails are unstable. The substitution of the double adiabatic model by evolution equations for the plasma pressures appears to suppress the instability in some cases and to allow the propagation of the solitary waves during long times.

\end{abstract}

%\pacs{52.27.Ny, 52.35.Sb, 52.38.-r, 52.65.-y}

\maketitle

\section{Introduction}

Solitary waves of various types are commonly observed in collisionless heliospheric plasmas.
A convincing observational evidence of large amplitude electromagnetic solitary wave propagating in the terrestrial environment was provided by Cluster multisatellite data near the magnetopause boundary \cite{Stasiewicz03a}. The  soliton, whose size is a few inertial lengths, is of slow type and is relatively stable as
\TPedit{it displayed very similar shapes when observed from two satellites at two different physical locations}.
 Other types of nonlinear waves in the form of fast magnetosonic shocklets are also observed with the Cluster satellites near the earth bow shock \cite{Stasiewicz03b}.
Compressive solitary structures or shocks are identified even further in the slow solar wind \cite{Perrone16}. Various structures in the form of single nonlinear Alfv\'en wave cycles, discontinuities, magnetic decreases, and shocks embedded in the turbulence of high-speed solar wind streams are reviewed in \cite{Tsurutani18}.
Magnetic humps or holes in total pressure balance, either in the form of isolated structures or in wave trains, are also commonly observed in planetary magnetosheath or in the solar wind \cite{Genot09}. They are often attributed to the saturation of the mirror instability. The latter being sub-critical, isolated magnetic holes can also be observed below the threshold of the mirror instability. Magnetic humps on the other hand, often require sufficiently large temperature anisotropy. These structures are clearly different from slow or fast modes as they are non-propagating in the plasma rest frame. Their propagation velocity is however difficult to measure precisely so that some uncertainty subsists in their identification. A complete determination of the various hydrodynamic as well as electromagnetic fields could permit to alleviate the ambiguity but this also remains a difficult observational task.
Even though these nonlinear structures are observed in \TPedit{almost collisionless} plasmas at scales of the order of a few ion Larmor radii, fluid modeling, possibly accounting for ion finite Larmor radius effects, appears to be sufficient to reproduce their main properties. Their amplitude is however large and an important challenge is to describe them as solutions of the fully nonlinear extended fluid equations rather than small amplitude asymptotic models.

Theoretical works on one-dimensional, localized, and travelling
structures have contributed to the understanding of the propagation
of nonlinear and low-frequency waves in plasmas. In the small
amplitude limit, these waves are governed by standard integrable
equations such as the Korteweg-de Vries (KdV) \cite{Kever:1969}, the
Modified KdV (MKDV) \cite{Kakutani:1969}, the Derivative Nonlinear
Schr\"odinger equation
(DNLS) \cite{Rogister:1971}, \cite{mjolhus:1976}, \cite{mjolhus:1988},
and the triple-degenerate DNLS equation \cite{Hada:1993}. Some of them admit
solitonic solutions and relations between their propagation velocity
and their amplitudes exist (see \cite{Mjolhus_06} and references
therein). \TPedit{These small amplitude asymptotic equations are also well suited to address questions related to perturbations of these solitary structures, such as e.g. the nontrivial effect of dissipation on Alfv\'en solitons \cite{Sanchez10},  or to the role played by non-maxwellian distribution functions on the shape and existence of solitons \cite{Mamun97, Tribeche12, Williams13}.} However, for finite amplitude, these localized structures
should be studied in a more general framework such as the
magnetohydrodynamic models extended to include dispersive and/or
dissipative effects. After assuming the travelling wave ansatz, the
system of partial differential equations becomes a set of ordinary
differential equations that can be used to investigate the existence
of solitary waves and discontinuities. This technique has been used
to study the structure of intermediate shock waves in the
resistive-magnetohydrodynamics (MHD) \cite{Hau_89}, the resistive
Hall-MHD \cite{Hau_90}, and Hall-MHD with a double-adiabatic
pressure tensor \cite{Sanchez:2013} systems, and also rotational
discontinuities in the Hall-MHD model with finite-Larmor-radius
(FLR) and scalar pressure \cite{Hau_1991}. Exact solitary waves
solutions in the Hall-MHD model for cold \cite{McKenzie:2001} and
warm plasmas with scalar \cite{McKenzie:2002} and double-adiabatic
pressure models \cite{Mjolhus_06} have been also found.

In the case of the Hall-MHD model with a double adiabatic pressure
tensor, the traveling wave ansatz leads to a pair of coupled
ordinary differential equations that governs the normalized
components of the magnetic field normal to the propagation
direction, named $b_y$ and $b_z$. Such a system has a hamiltonian
structure and is reversible, i.e. solutions are invariant under the
transformation $(\zeta, b_y, b_z\rightarrow -\zeta, -b_y, b_z)$,
with $\zeta$ the independent variable. Adding FLR effects does not
change the reversible character of the dynamical system but it
increases the effective dimension from two to four
\cite{Mjolhus_2009}. Numerical evidence about the existence of
solitary waves in the parametric domain where the upstream state is
a saddle-center was also given \cite{Mjolhus_2009}. The hamiltonian
character of the dynamical system with FLR effects is an open and
interesting topic, especially because an energy conservation theorem
is not known for the Hall-MHD model with double adiabatic pressure
and without FLR effects.

This work investigates the existence and stability of solitary waves
in the FLR-Hall-MHD model with double adiabatic pressure tensor.
Section \ref{Sec:Model} follows Ref. \cite{Mjolhus_2009} closely,
and presents in a concise way the procedure to find the dynamical
system that governs the solitary waves. The details of the method
are given in Appendix \ref{Sec:Dynamical:System}, where few
discrepancies with the results of Ref. \cite{Mjolhus_2009} are
highlighted. Section \ref{Sec:Model} also discusses the main
properties of the dynamical system and takes  advantage of  some
geometrical arguments related with the dimension, reversible
character, and the stability of the upstream state of the solitary
waves, to anticipate the organization of the solitary waves in
parameter space. Such organization, which was briefly suggested in
Ref. \cite{Sanchez:2013}, lies on well-known results for homoclinic
orbits in reversible systems \cite{Champneys_98}. Section
\ref{Sec:Existence} \ESedit{introduces} a numerical procedure that proves
rigorously the existence of solitary waves and uses it to compute
them in several parametric regimes. The stability of the solitary
waves is investigated in Sec. \ref{Sec:Stability}, where the
solutions of the dynamical system are introduced as initial
conditions in the FLR-Hall-MHD model. \GSAedit{Two different closure models are considered in the simulations: the already mentioned double adiabatic model, which helped us to keep low the dimension of the dynamical system in the analysis of the existence of solitary waves, and also a dynamic model for the pressure evolution that yields an energy-conserving system. The physical
consequences of these results for the existence and stability of
solitary waves in collisionless plasmas and the role played by the
propagation angle with respect to the ambient magnetic field and the
wave velocity are highlighted throughout Sections
\ref{Sec:Existence} and \ref{Sec:Stability}}. Section
\ref{Sec:Conclusions} summarizes the conclusions of the work.

\section{The FLR-Hall MHD model\label{Sec:Model}}

The analysis is carried out in the framework of the FLR-Hall-MHD
system. Mass density $\rho$, plasma (i.e. ion) flow velocity
$\mathbf{v}$, and magnetic field $\mathbf{B}$ are governed by
\begin{equation}
\frac{\partial \rho}{\partial t} + \nabla \cdot \left( \rho
\mathbf{v} \right) = 0 \label{e:continuity_FLR}
\end{equation}
\begin{equation}
\frac{\partial}{\partial t} \left( \rho \mathbf{v} \right) + \nabla
\cdot \left[ \rho \mathbf{v}\mathbf{v} + \bar{\mathbf{P}}_i +p_e
\hspace{1mm} \bar{\mathbf{I}} + \frac{1}{4\pi} \left( \frac{1}{2}
B^2 \bar{\mathbf{I}} - \mathbf{B} \mathbf{B} \right) \right] = 0
\label{e:momentum_FLR}
\end{equation}
\begin{equation}
\frac{\partial \mathbf{B}}{\partial t} = \nabla \times \left[
\mathbf{v} \times \mathbf{B} - \frac{m_i c}{4 \pi e \rho}
\left(\nabla \times \mathbf{B}\right) \times \mathbf{B}\right]
\label{e:magnetic_induction_FLR}
\end{equation}
where $m_i$ is the ion mass, $c$ the speed of light,  $e$ the
electron charge, and $\bar{\mathbf{I}}$ the identity tensor. We
assumed that the electron pressure is isotropic and follows an
isothermal equation of state $p_e = \rho v_{se}^2$ with $v_{se}^2 =
k_B T_e/m_i$ the electron contribution to the ion-acoustic velocity,
$T_e$ the electron temperature and $k_B$ Boltzmann's constant. The
ion pressure tensor $\bar{\mathbf{P}}_i$ is written as
\begin{equation}
\bar{\mathbf{P}}_i = \bar{\mathbf{P}}_{i}^{\left(0\right)} +
\bar{\mathbf{P}}_{i,1}^{\left(1\right)}+\bar{\mathbf{P}}_{i,2}^{\left(1\right)}+\bar{\mathbf{P}}_{i,3}^{\left(1\right)}
\label{Eq:Ion:Pressure}
\end{equation}
where the tensor with superscript $0$ represents the gyrotropic
contribution and reads
\begin{equation}
 \label{e:Pi0}
\bar{\mathbf{P}}_{i}^{\left(0\right)}= p_{\parallel} \mathbf{e}_b
\mathbf{e}_b + p_{\perp} \left(\bar{\mathbf{I}} - \mathbf{e}_b
\mathbf{e}_b\right)\equiv\bar{\mathbf{P}}_{i,
\parallel}^{\left(0\right)}+\bar{\mathbf{P}}_{i,
\perp}^{\left(0\right)}
\end{equation}
with $p_{\parallel}$ and $p_\perp$ the parallel and perpendicular
pressures and $\mathbf{e}_b=\mathbf{B}/B$ the unit vector along the
magnetic field. Tensors with superscript $1$ in Eq.
\eqref{Eq:Ion:Pressure} represents the FLR corrections and are
given by \cite{Macmahon:1965,Yajima:1966}
\begin{equation}
\label{e:Pi11} \bar{\mathbf{P}}_{i,1}^{\left(1\right)} =
\frac{1}{\Omega_{ci}} \left[ \frac{1}{4} \mathbf{e}_b \times \left(
\nabla \mathbf{v} + \nabla \mathbf{v}^T \right) \cdot
\bar{\mathbf{P}}_{i, \perp}^{\left(0\right)} + \textmd{transp.}
\right]
\end{equation}
\begin{equation}
\label{e:Pi21} \bar{\mathbf{P}}_{i,2}^{\left(1\right)} =
-\frac{1}{\Omega_{ci}} \left[ \mathbf{e}_b \left( \nabla \times
\mathbf{v} \right) \cdot
\bar{\mathbf{P}}_{\GSAedit{i,}\perp}^{\left(0\right)} +
\textmd{transp.} \right]
\end{equation}
\begin{equation}
\label{e:Pi31} \bar{\mathbf{P}}_{i,3}^{\left(1\right)} =
\frac{2}{\Omega_{ci}} \left[ \mathbf{e}_b \left(
\bar{\mathbf{P}}_{i,\parallel}^{\left(0\right)} \cdot \nabla \right)
\times \mathbf{v} + \textmd{transp.} \right]
\end{equation}
where $\Omega_{ci} = eB/\left(m_i c\right)$ is the local ion gyro
frequency and the notation \emph{+transp} means that one should sum
the transpose of the tensor immediately to the left in the square
bracket. The equations are completed  with the following
double-adiabatic model for the equations of state
\begin{align}
\frac{p_{\parallel} B^2}{\rho^3} =& \textmd{const.}
\label{e:p_par_Hall}\\
 \frac{p_{\perp}}{\rho B} =& \textmd{const.}
\label{e:p_perp_Hall}
\end{align}
Hereafter, subscript $0$ will be used to denote the unperturbed
variables. Therefore, $\rho_0$, $B_0$, $p_{\parallel 0}$ and
$p_{\perp 0}$ correspond to the values of $\rho$, $B$,
$p_{\parallel}$ and $p_{\perp}$ upstream from the solitary wave. We
also now introduce a cartesian frame of reference with the $x$-axis
along the propagation direction of the wave, and the $y$- and
$z$-axis chosen such that the upstream magnetic field has no
component in the $y$-direction. Such a frame is linked to the
solitary wave and moves at velocity $v_{x0}$ with respect to the
unperturbed plasma. In the upstream region, i.e. at $x\rightarrow +\infty$,
plasma velocity and magnetic field then read
\begin{align}
\mathbf{v}(x\rightarrow +\infty) =\ v_{x0} \mathbf{e}_x\\
\mathbf{B}_0 =\ B_0\left(\cos\theta \mathbf{e}_x + \sin\theta
\mathbf{e}_z\right)
\end{align}
with $\mathbf{e}_x$, $\mathbf{e}_y$, and $\mathbf{e}_z$  unit
vectors along the axes of the cartesian frame. Therefore, the
solitary wave propagates along the positive (negative) $x$ direction
for $v_{x0}<0$ ($v_{x0}>0$). We will consider the case $v_{x0}<0$
and will use the wave velocity $C=-v_{x0}$.

If the analysis is restricted to stationary ($\partial/\partial t=0$)
and one-dimensional waves ($\partial/\partial y$=$\partial/\partial
z$=0), then one finds that $B_x$ is constant ($B_x= B_0\cos\theta)$
and the FLR-Hall-MHD model becomes the following set  of ordinary
differential equations
\begin{equation}
\frac{d\bm{\xi}}{d\hat{x}} =
\mathbf{f}\left(\bm{\xi}\right).\label{Eq:Dynamical:System}
\end{equation}
The state vector of this dynamical system is five-dimensional,
$\bm{\xi} = \left[u_x\ \ u_y\ \ u_z\ \ b_y\ \ b_z \right]^T$, and it
involves the normalized velocity
$\mathbf{u}=u_x\mathbf{e}_x+u_y\mathbf{e}_y+u_z\mathbf{e}_z$ and
magnetic field components
$\mathbf{b}=b_y\mathbf{e}_y+b_z\mathbf{e}_z$ with $\mathbf{u}\equiv
\mathbf{v}/v_{x0}$ and $b_{y,z}\equiv B_{y,z}/B_0\sin\theta$. The
independent variable in Eq. \eqref{Eq:Dynamical:System} is the
normalized length $\hat{x} = x/\ell$, with
\begin{equation}
\ell = \frac{v_A^2\cos\theta}{\Omega_{ci,0}v_{x0}},
\end{equation}
$v_A=\sqrt{B_0^2/4\pi\rho_0}$ the Alfv\'en velocity, and
$\Omega_{ci,0} = eB_0/\left(m_i c\right)$ the upstream ion cyclotron
frequency. In Ref. \cite{Mjolhus_2009}, a $\cos\theta$ factor was
missed. The dynamical system involves five parameters:
\begin{align}
\theta, \ \  M_A = \frac{v_{A}^2}{v_{x0}^2},\ \ M_e =
\frac{v_{se}^2}{v_{x0}^2}, \ \ M_i = \frac{v_{\perp}^2}{v_{x0}^2},\
\
 a_p = \frac{p_{\parallel 0}}{p_{\perp
0} },\label{Eq:Parameters}
\end{align}
\TPedit{ $\theta$ being} the angle between the propagation direction and the
ambient unperturbed magnetic field and
$v_\perp^2=p_{\perp_0}/\rho_0$. This work investigates the effect of
$C/V_A$ and $\theta$ on the properties of the solitary waves and
will fix the other parameters according to the two cases shown in
Table \ref{Table:Par}. The explicit form of the vector flow
$\mathbf{f}$ in Eq. \eqref{Eq:Dynamical:System} and a comparison
with the results of Ref. \cite{Mjolhus_2009} are provided in
Appendix \ref{Sec:Dynamical:System}.

 \begin{table}[h]
 \caption{Solitary waves parameters.\label{Table:Par}}
 \begin{ruledtabular}
 \begin{tabular}{c c c  c}
 Case &  $a_p$ & $v_\perp/v_A$ & $v_{se}/v_\perp$\\
\hline
1     &  $1$     & $0.4$        & $1$\\
2     &  $1.5$   & $1.2$        & $0.3$\\
 \end{tabular}
 \end{ruledtabular}
 \end{table}

\subsection{Properties of the FLR-Hall MHD dynamical system}

Before discussing interesting physical features of the solitary
waves in Sec. \ref{Sec:Existence}, we now summarize some purely
mathematical results that are \GSAedit{essential} \ESedit{in order} to organize the
waves in the parameter space and design numerical algorithms to
compute them and prove their existence rigorously. An important
property of Eq. \eqref{Eq:Dynamical:System} is the existence of a
manifold $U$  that orbits cannot cross. As shown in Appendix
\ref{Sec:Singularity}, Eq. \eqref{Eq:Dynamical:System} is singular
for the manifold determined by the condition
\begin{equation}
\Gamma_R(\bm{\xi}) = 0,
\end{equation}
with $\Gamma_R$ given by Eq. \eqref{Eq:Gamma}. The role of this set
is similar to the sonic circle found in the Hall-MHD model
\cite{Mjolhus_2009}.

We first note that the upstream state $\bm{\xi}_0 = \left[1\ \ 0\ \
0\ \ 0\ \ 1 \right]^T$ is an equilibrium state of Eq.
\eqref{Eq:Dynamical:System} because it satisfies
$\mathbf{f}(\bm{\xi}_0)=0$. Another interesting element is the
stable (unstable) manifold $W^s$ ($W^u$) of $\bm{\xi}_0$, which is
the set of forward (backward) in $\hat{x}$ trajectories that
terminate at $\bm{\xi}_0$. Since solitary waves are localized
structures that approach upstream and downstream to $\bm{\xi}_0$,
i.e. $\bm{\xi}\rightarrow \bm{\xi}_0$ as $\hat{x}\rightarrow \pm
\infty$, these special solutions belong to the intersections of the
stable and the unstable manifolds of $\bm{\xi}_0$. They are called
homoclinic orbits. As explained below, their organization in
parameter space depends on (i) the dimension of the phase space,
(ii) the occurrence of symmetries, and (iii) the dimensions of the
stable and the unstable manifolds of $\bm{\xi}_0$. These three
topics are discussed below.

In the particular case of Eq. \eqref{Eq:Dynamical:System}, the
dimension of the phase space, given by the number of components of
$\bm{\xi}$, is five. However, as shown in Appendix
\ref{Sec:Dynamical:System}, there is a function $H(\bm{\xi})$ that
satisfies $dH/d\hat{x}=0$ \cite{Hau_1991,Mjolhus_2009}, i.e. it is
conserved. As a consequence, the effective dimension of our system
is four. Regarding symmetries, one readily verifies that Eq.
\eqref{Eq:Dynamical:System} is reversible because it admits the
involution $G\mathbf{f}(\bm{\xi}) = -\mathbf{f}(G \bm{\xi})$ with
\begin{equation}
G: \left(u_x,u_y,u_z,b_y,b_z\right)\rightarrow
\left(u_x,-u_y,u_z,-b_y,b_z\right)\label{Eq:Involution}
\end{equation}
The subspace $S: u_y=b_y=0$, a key element for the later computation
of the solitary waves, is called the symmetric section of the
reversibility. Interestingly, a symmetric solitary wave exists if
the unstable manifold of $\bm{\xi}_0$ intersects the symmetric
section at a given point. The reason is that, by the reversibility
property, $W^s$ should also intersect $S$ at the same point.

\begin{figure}[h]
    \centering
    \noindent\includegraphics[scale=1]{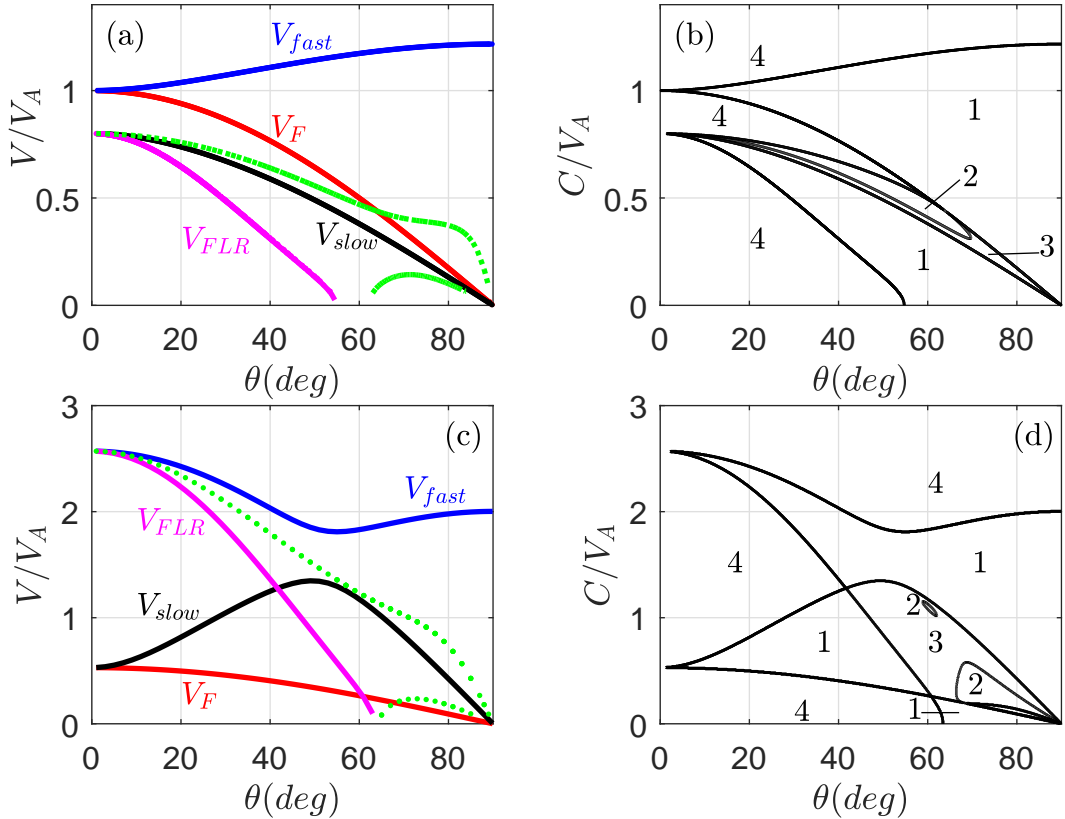}
    \caption{Characteristic velocities (left) and domains of stability of $\bm{\xi}_0$ in the $C/V_A-\theta$ plane (right). The regions are  (1) Saddle-Center, (2) Focus-Focus, (3) Saddle-Saddle, (4) Center-Center.
The parameters used for panels (a) and (b) [(c) and (d)] correspond
to case 1 (2) in Table \ref{Table:Par}. The green dotted lines in
panels (a) and (c) are the velocities making $p_1=0$ in Eq.
\eqref{Eq:biquadratic}.
     }
    \label{Fig:Domains}
\end{figure}

On the other hand, the tangent spaces of $W^s$ and $W^u$ have the
same dimensions as the stable and unstable spaces of the
linearization of $\mathbf{f}$ at $\bm{\xi}_0$. Substituting
$\bm{\xi} = \bm{\xi}_0+\bm{\xi}_1$ in Eq.
\eqref{Eq:Dynamical:System} with $\bm{\xi}_1$ a small perturbation
and dropping higher order terms yield
\begin{equation}
\frac{d\bm{\xi}_1}{d\hat{x}} \approx
\bar{\bm{J}}\mid_{\bm{\xi}_0}\bm{\xi}_1
\end{equation}
where $\bar{\bm{J}}\mid_{\bm{\xi}_0}$ is the Jacobian matrix of
$\mathbf{f}$ at $\bm{\xi}_0$. If we now assume $\bm{\xi}_1(\hat{x})
= \hat{\bm{\xi}}_1e^{\lambda \hat{x}}$, the condition for nontrivial
$\hat{\bm{\xi}}_1$ is
\GSAedit{$\det(\bar{\bm{J}}\mid_{\bm{\xi}_0}-\lambda
\bar{\bm{I}})=0$}. Such a condition gives $\lambda = 0$, which is a
consequence of the invariant $H$, and the following characteristic
equation with a biquadratic structure that reflects the involution
given by Eq. \eqref{Eq:Involution}
\begin{equation}
p_2\lambda^4 + p_1\lambda^2+p_0=0\label{Eq:biquadratic},
\end{equation}
where $p_2$, $p_1$ and $p_0$ are certain constants that just depend
on the five parameters of Eq. \eqref{Eq:Parameters} (find their
explicit forms in Ref. \cite{Mjolhus_2009}). These coefficients
contain important information that will help us to connect the
mathematical results with the physics of the solitary waves.
Coefficient $p_0$ vanishes when the propagation velocity $C$
coincides with one of the non-dispersive MHD velocities, i.e. the
system obtained after neglecting the Hall and the FLR terms. The MHD
velocities are the fast ($V_{fast}$) and the slow ($V_{slow}$)
magnetosonic velocities, and the firehose velocity ($V_F$), which
reduces to the intermediate or shear Alfv\'en velocity in the case
of isotropic pressure. Coefficient $p_1$ and $p_2$ vanish when the
propagation velocity $C$ is equal to the acoustic velocity corrected
with FLR effects ($V_s$) and the velocity $V_{FLR}$ defined by the
condition $\Gamma_R(\xi_0)=0$, respectively \cite{Mjolhus_2009}. For
the parameters of Table \ref{Table:Par}, these velocities are
plotted versus the propagation angle in panels (a) and (c) of Fig.
\ref{Fig:Domains}.

The generic cases of the solutions of Eq. \eqref{Eq:biquadratic}
are: (i) saddle-center, $\lambda_{1,2} = \pm \kappa$ and
$\lambda_{3,4}=\pm i\omega$, (ii) saddle-saddle, $\lambda_{1,2} =
\pm \kappa_1$ and $\lambda_{3,4} = \pm \kappa_2$, (iii) focus-focus,
$\lambda_{1,2} = \kappa\pm i \omega$ and $\lambda_{3,4} = -\kappa\pm
i \omega$\GSAedit{, and}  (iv) center-center, $\lambda_{1,2}=\pm
i\omega_1$ and $\lambda_{3,4}=\pm i\omega_2$. Panel (b) and (d) in
Fig. \ref{Fig:Domains} shows the domains of stability of
$\bm{\xi}_0$ in the $C/V_A-\theta$ plane for cases (1) and (2) in
Table \ref{Table:Par}. Although this set of parameters yields to
unstable solitary waves (see Sec. \ref{Sec:Stability}), they have
been used throughout this work because they were used in Ref.
\cite{Mjolhus_2009}. Working with the same physical parameters eases
the comparison of our results and highlights the main novelties
related with the existence of the solitary waves and their
organization in parameter space. This particular case is also
illustrative because, as shown in Fig. \ref{Fig:Domains} the four
stability regions of $\bm{\xi}_0$, exist in the $C/V_A-\theta$
plane.

Taking into account that the effective dimension of the system is
four and its reversible character, well-known theoretical results on
the existence of homoclinic orbits can be directly applied to our
case (find a review in Ref. \cite{Champneys_98}). To fix ideas,
consider the situation with given $M_e$, $M_i$, and $a_p$ values and
let us discuss the organization of solitary waves in the
$M_A-\theta$ plane (as already done in Ref. \cite{Mjolhus_2009}).
Unless very specific resonance conditions are fulfilled, no solitary
wave occurs when $\bm{\xi}_0$ is a center-center because such a
point has no stable or unstable manifold and orbits cannot connect
with it. For values of $M_A$ and $\theta$ making $\bm{\xi}_0$ a
saddle-center, the stable and unstable manifolds have dimension
equal to one, and an homoclinic orbit exists \ESedit{if the two coincide,}
$W^s=W^u$. In general, the intersection of the one-dimensional
manifold $W^u$ with the two-dimensional symmetric section is
expected to occur for specific parameter values that form branches
in the $M_A-\theta$ plane. For parameter values where $\bm{\xi}_0$
is hyperbolic, i.e. saddle-saddle and focus-focus, $W^u$ has
dimension two. The intersection of such a two-dimensional manifold
with the two-dimensional symmetric section in a four-dimensional
phase space is generic and solitary waves are expected to exist in
continuous regions in the $M_A-\theta$ plane. This is called a
continuous spectrum. According to this discussion, we expect that
\GSAedit{in panels (b) and (d)} of Fig. \ref{Fig:Domains} we will
find branches of solutions in region 1, a continuous spectrum in
regions 2 and 3, and no solitary wave in region 4.

\section{FLR-Hall-MHD solitary waves\label{Sec:Existence}}

\subsection{Saddle-Center Domain}

According to previous geometrical arguments, solitary waves are
organized in branches within the saddle-center domain. These
branches can be computed, and their existence proved rigorously, by
using the following bisection algorithm (see details in Ref.
\cite{Sanchez:2017}). For a given set of parameters, Eq.
\eqref{Eq:Dynamical:System} is integrated with initial condition
equal to
\begin{equation}
\bm{\xi}(\hat{x}=0) = \bm{\xi}_0+ \epsilon \bm{\xi}^u,
\end{equation}
where $\epsilon$ is a small parameter ($10^{-9}$ in our
calculations) and $\bm{\xi}^u$ is the unstable eigenvector of
$\bar{J}\mid_{\bm{\xi}_0}$, i.e. the eigenvector with the positive
 and real eigenvalue. Such initial condition
guarantees that the orbits leaves the equilibrium state from the
linearization of the unstable manifold. The orbit is computed up to
the value $\hat{x}^*$ that satisfies the condition
$b_y(\hat{x}^*)=0$. We then recorded the value of $u_y(\hat{x}^*)$.
Such a procedure is repeated by covering a range of $C/V_A$ (or
$\theta$) values, and we then construct a diagram with
$u_y(\hat{x}^*)$ versus $C/V_A$ (or $\theta$). Each time a change of
sign in $u_y(\hat{x}^*)$ occurs, it means that $u_y(\hat{x}^*)$
passes through zero and there is an orbit leaving the unstable
manifold of $\bm{\xi}_0$ and hitting the symmetric section.
Therefore, a solitary wave exists.

\begin{figure}[h]
    \centering
    \noindent\includegraphics[scale=.9]{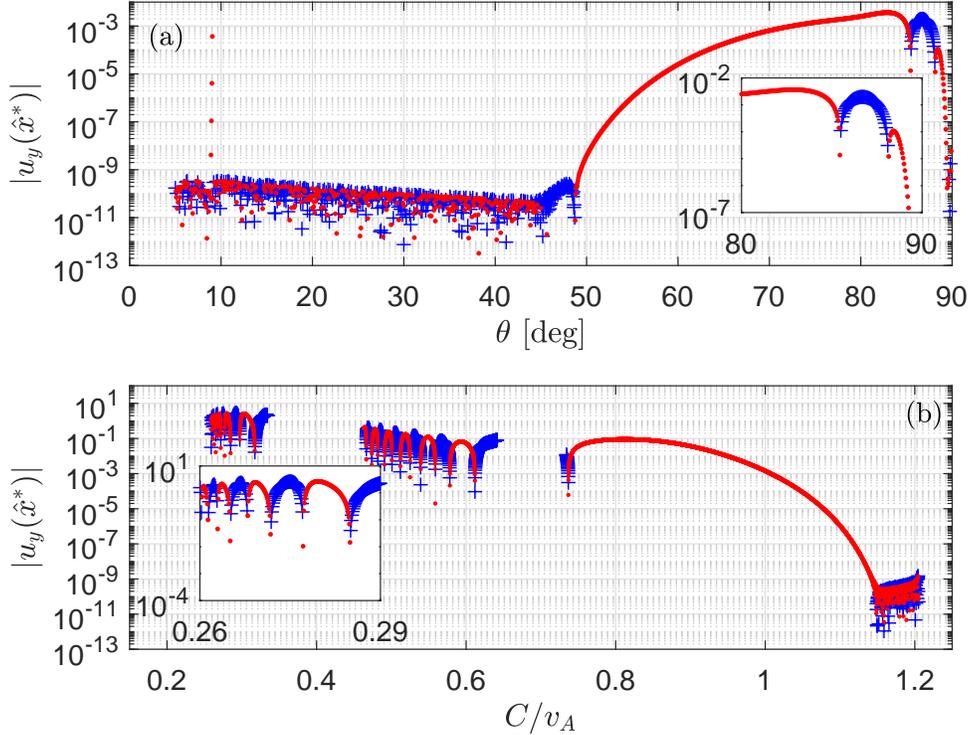}
    \caption{Panels (a) and (b) show  $u_y(\hat{x}^*)$ versus $\theta$ for $C/V_A = 1$, and $u_y(\hat{x}^*)$ versus $C/V_A$ for $\theta= 75^\circ$, respectively. Other parameters correspond to case 1 in Table \ref{Table:Par}. Positive (negative) values of $u_y(\hat{x}^*)$ are denote with blue crosses (red dots).}
    \label{Fig:Sign:Change}
\end{figure}

Panel (a) in Fig. \ref{Fig:Sign:Change} shows  $u_y(\hat{x}^*)$
versus $\theta$ for $C/V_A = 1$ and the \GSAedit{parameters} of case
1. The zeros of $u_y(\hat{x}^*)$ have been highlighted by plotting
the absolute value of $u_y(\hat{x}^*)$ in logarithmic scale and
using blue crosses and red dots for positive and negative values of
$u_y(\hat{x}^*)$, respectively. Clearly, solitary waves exist for
$\theta \approx 88.15^\circ $ and $\theta = 85.45^\circ $ (see
inset) and $\theta = 48.7^\circ$. For $\theta< 47^\circ$, where the
crosses and dots are mixed and do not follow a smooth curve, we
cannot guarantee (neither rule out) the existence of branches. The
reason is that the values of $u_y(\hat{x}^*)$ are very small and
they fall below our numerical error, which \ESedit{is a combination of factors including} the finite value
of $\epsilon$, the integration error, and the finite precision \ESedit{arithmetic (double precision floating-point format used here)}.

\begin{figure}[h]
    \centering
    \noindent\includegraphics[scale=.9]{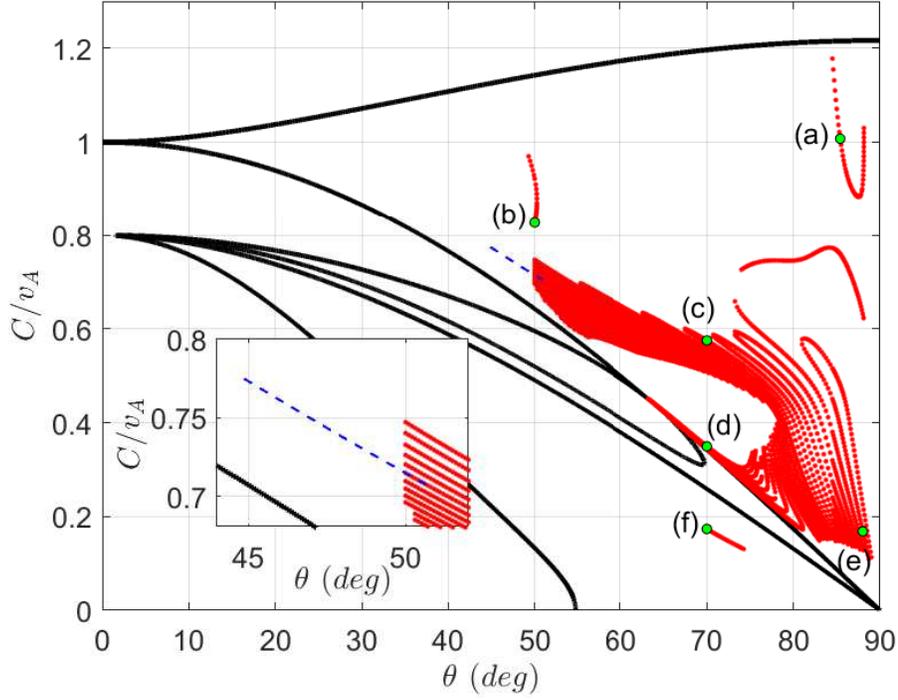}
    \caption{Branches of solitary waves in the $C/V_A-\theta$ plane for case 1.}
    \label{Fig:Branches}
\end{figure}

Panel (b) in Fig. \ref{Fig:Sign:Change} shows a similar diagram but
varying $C/v_A$ for $\theta=75^\circ$. For  $C/v_A>1.15$ we find
again a parametric region where  $u_y(\hat{x}^*)$ is smaller than
our error. For lower propagation velocity, as $C/v_A$ decreases, one
first finds a wave with $C/v_A\approx 0.73$, a gap, a velocity range
with many waves, a second gap, and another region with several waves
(see inset). The gaps appear because, when launching an integration
along the unstable manifold, the orbit hit the singular  manifold
$\Gamma_R(\bm{\xi})=0$. An intensive parametric survey constructing
diagrams such as the ones in Fig. \ref{Fig:Sign:Change}, allowed us to
present the branches of solitary waves in the two saddle-center
regions \GSAedit{of the} $C/v_A-\theta$ plane (see Fig.
\ref{Fig:Branches}). In the saddle-center region \ESedit{delimited} by the
firehose and the fast magnetosonic velocities there are many
branches of solutions, specially close to $V_F$. For large
propagation angles, it is even possible to find solitary waves with
propagation velocities larger than the Alfv\'en velocity.

\ESedit{The limiting factor in overcoming the numerical problems in 
the calculations presented here is the use of finite precision arithmetic.
This is indicated by the fact that further reducing  $\epsilon$ 
or increasing the accuracy of the integration does not lead to resolution of  
solitary wave branches for $\theta< 50^\circ$ in \reffig{Fig:Branches}. 
In order to progress further we implemented our method in the computer algebra system Mathematica,
taking advantage of its arbitrary precision capabilities. Using 30 digits of working precision,
error tolerance of 20 digits in the integrator and taking $\epsilon=10^{-15}$ allows us to resolve branches 
of solitary waves in this problematic regime. We show one example branch as a dashed line in Fig.~\ref{Fig:Branches}.
This branch is tracked until numerical errors once again prevent us from isolating solitary wave solutions.
Even though a further increase in precision could help proceed towards lower values of $\theta$, 
the computations quickly become very expensive and
we do not pursue an exhaustive determination of branches.
The main point we illustrate here is that the difficulties in locating solitary waves for smaller angles are indeed numerical 
and can be overcome by increasing the precision of the calculations.
}

\ESedit{We note that works on electromagnetic solitary waves in
relativistic plasma (laser-plasma interaction framework) have encountered similar difficulties.
In particular, } 
waves in regions such as the one shown 
in Fig.~\ref{Fig:Sign:Change}(a) with
$\theta< 47^\circ$, where the residual value of $u_y(\hat{x}^*)$ is
very small, were erroneously taken as true waves with a continuous
spectrum in \ESedit{early} works. It was
later shown that they should be organized in branches in the
saddle-center domain and the claimed waves were numerical artifacts, 
\ESedit{see \cite{Sanchez:2017} and references therein}. Similarly, the FLR-Hall-MHD solitary wave
presented in Fig. 2 of Ref. \cite{Mjolhus_2009} is not a true
localized structure because: (i) the values of $\theta$ and $C/V_A$
were selected without looking for a branch (a relation between
$\theta$ and $C/V_A$) and (ii) the author found a value of
$u_y(\hat{x}^*)\approx 10^{-7}$ within this (numerically difficult)
parametric domain.

Figure \ref{Fig:Branches} \GSAedit{also} shows that many solitary
waves can exist with propagation velocity covering a broad range
between the sonic and the fast magnetosonic velocities (and not only
close to $V_{fast}$ as concluded in Ref. \cite{Mjolhus_2009}). In
the Hall-MHD model, these waves are of type dark and were termed the
fast magnetosonic family \cite{Mjolhus_06}. As shown below, the
solitary waves found in the FLR-Hall-MHD model are also dark for
that regime. Above point (d) there is a blank
region because the orbits started along the unstable manifold hit
the surface $\Gamma_R = 0$.

\begin{figure}[h]
    \centering
    \noindent\includegraphics[scale=.9]{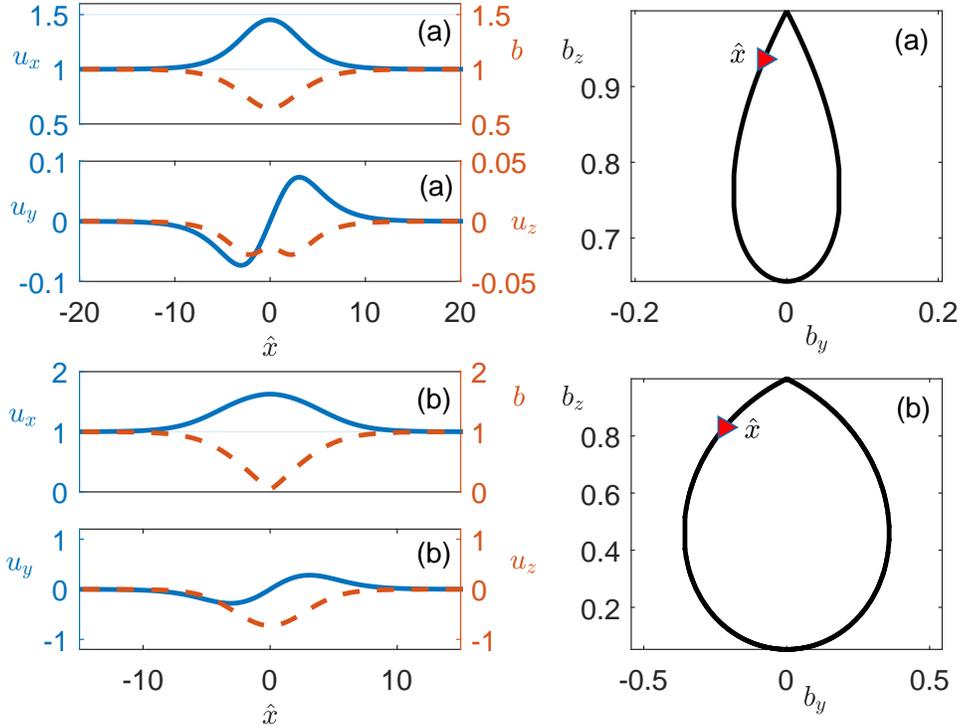}
    \caption{Solitary waves named (a) and (b) in Fig. \ref{Fig:Branches}.}
    \label{Fig:SC:Orbit:1}
\end{figure}

\begin{figure}[h]
    \centering
    \noindent\includegraphics[scale=.9]{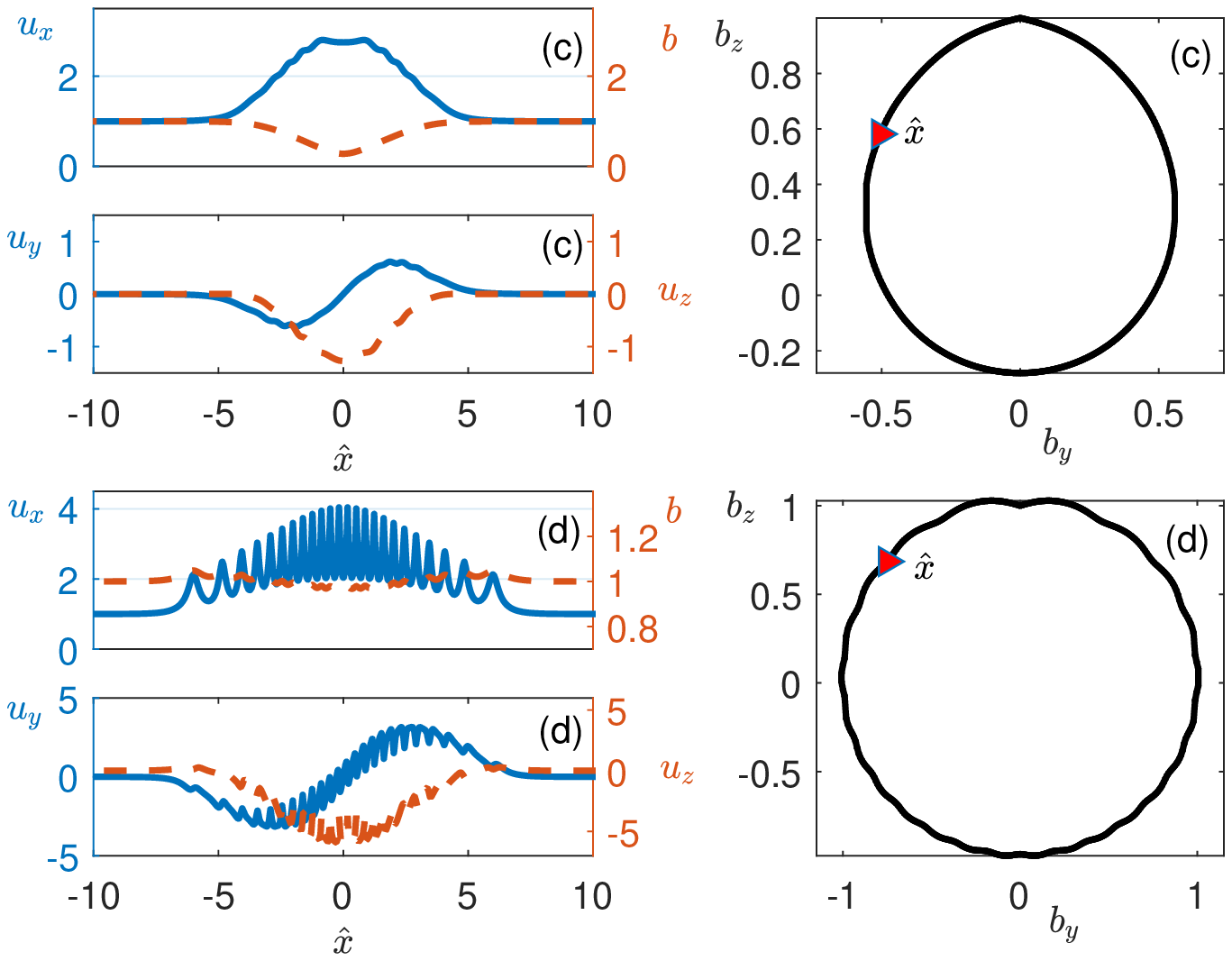}
    \caption{Solitary waves named \EBedit{(c) and (d)} in Fig. \ref{Fig:Branches}.}
    \label{Fig:SC:Orbit:2}
\end{figure}

\begin{figure}[h]
    \centering
    \noindent\includegraphics[scale=.9]{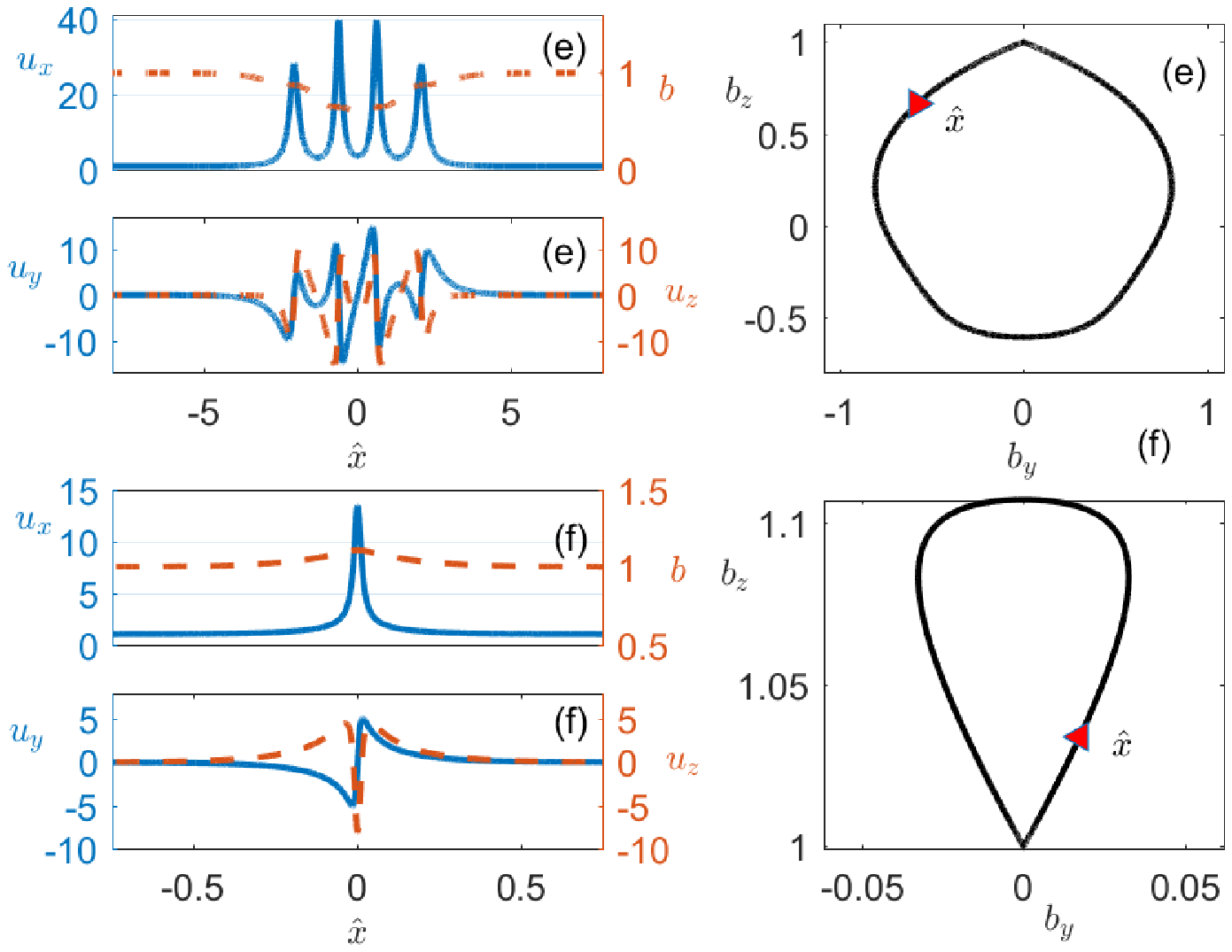}
    \caption{Solitary waves named (e) and (f) in Fig. \ref{Fig:Branches}.}
    \label{Fig:SC:Orbit:3}
\end{figure}

In order to illustrate the different types of solitary waves, we
selected six cases in Fig. \ref{Fig:Branches} and labeled them with
letters from (a) to (e). For all of them, we plotted the velocity
components, the modulus of the magnetic field normal to the
propagation direction $b$, and the magnetic hodograph (see Figs.
\ref{Fig:SC:Orbit:1}-\ref{Fig:SC:Orbit:3}). In the latter, we
denoted with an arrow the increasing direction of $\hat{x}$ and kept
the same scale for both axes to ease the interpretation of the wave
polarization. Since all the selected waves have $u_x>1$ at
$\hat{x}=0$, the relation $\rho/\rho_0=1/u_x$ indicates that the
densities at the center of the structures are lower than the
background value.

Solitary waves (a)-(e) belong to the saddle-center domain with
propagation velocities between the firehose  and the fast
magnetosonic velocities. Waves (a) and (b) are dark solitary waves,
i.e. the magnetic field exhibits a minimum at the center of the
structure (see Fig. \ref{Fig:SC:Orbit:1}). Wave (a), which has a
larger propagation angle and velocity, exhibits a much lower
depression of the magnetic field and its polarization is more linear
as compared with wave (b). Waves (c), (d) and (e), showed in Figs
\ref{Fig:SC:Orbit:2} and \ref{Fig:SC:Orbit:3}, have been selected to
illustrate the set of branches that populate the \GSAedit{central}
region of Fig. \ref{Fig:Branches}. For a given propagation angle,
for instance $\theta = 70^\circ$ in cases (c) and (d), the solitary
waves develop more and more oscillations as the velocity decreases.
The polarization is almost circular. Wave (e), which propagates
almost normal to the ambient magnetic field, \GSAedit{has} magnetic
field \GSAedit{variations} of order unity, but much stronger changes
on the normalized velocity components (up to forty times the
propagation velocity). The central core of the solitary wave is also
complex and involves several peaks.

The saddle-center domain enclosed by the slow magnetosonic and the
FLR velocities is particularly interesting \GSAedit{from a physical
point of view}. No wave was found numerically in Ref.
\cite{Mjolhus_2009} for this domain with the FLR-Hall-MHD model.
Moreover, no solitary wave exists in the Hall-MHD model because the
upstream state is of type center. As shown in Fig.
\ref{Fig:Branches}, the FLR effect open new possibilities because a
branch of solutions occurs with propagation angle between
$70^\circ\le \theta \le 74^\circ$. The solitary wave named (f),
which is an example of such a branch, shows that for this domain the
solitary waves are of type bright, i.e. they exhibit a maximum of
the magnetic field at its center. This slow family presents small
(large) modulations of the magnetic (velocity) field components
(note the different scale of the left and right axes in the
$u_x-\hat{x}$ and $b-\hat{x}$ diagrams).

\subsection{Saddle-Saddle and Focus-Focus Domains}

In the saddle-saddle and focus-focus cases, since the dimension of
the unstable manifold is two and the solitary waves have a
continuous spectrum, the algorithm should be modified slightly. For
given parameter values, the initial condition in the saddle-saddle
and focus-focus cases are
\begin{align}
\bm{\xi}(\hat{x}=0) =& \bm{\xi}_0+ \epsilon\left(\cos\varphi
\bm{\xi}^u_1+\sin\varphi \bm{\xi}^u_2\right)\\
\bm{\xi}(\hat{x}=0) =& \bm{\xi}_0+ \epsilon
Re\left(e^{i\varphi}\bm{\xi}^{u}\right)\label{Eq:IC:Focus:Focus}
\end{align}
with $\bm{\xi}^u_1$ and $\bm{\xi}^u_2$ the eigenvectors with
positive eigenvalues for the saddle-saddle case, and
$\bm{\xi}^{u}$ any of the two eigenvectors with eigenvalues having
a positive real part for the focus-focus case. Angle $\varphi$ is a
numerical parameter that controls the position of the initial
condition in the linearization of the unstable manifold. The
numerical scheme is similar to the saddle-center case, but now we
need to look for the change of sign of $u_y(\hat{x}^s)$ as  a
function of $\varphi$.

\begin{figure}[h]
    \centering
    \noindent\includegraphics[scale=.9]{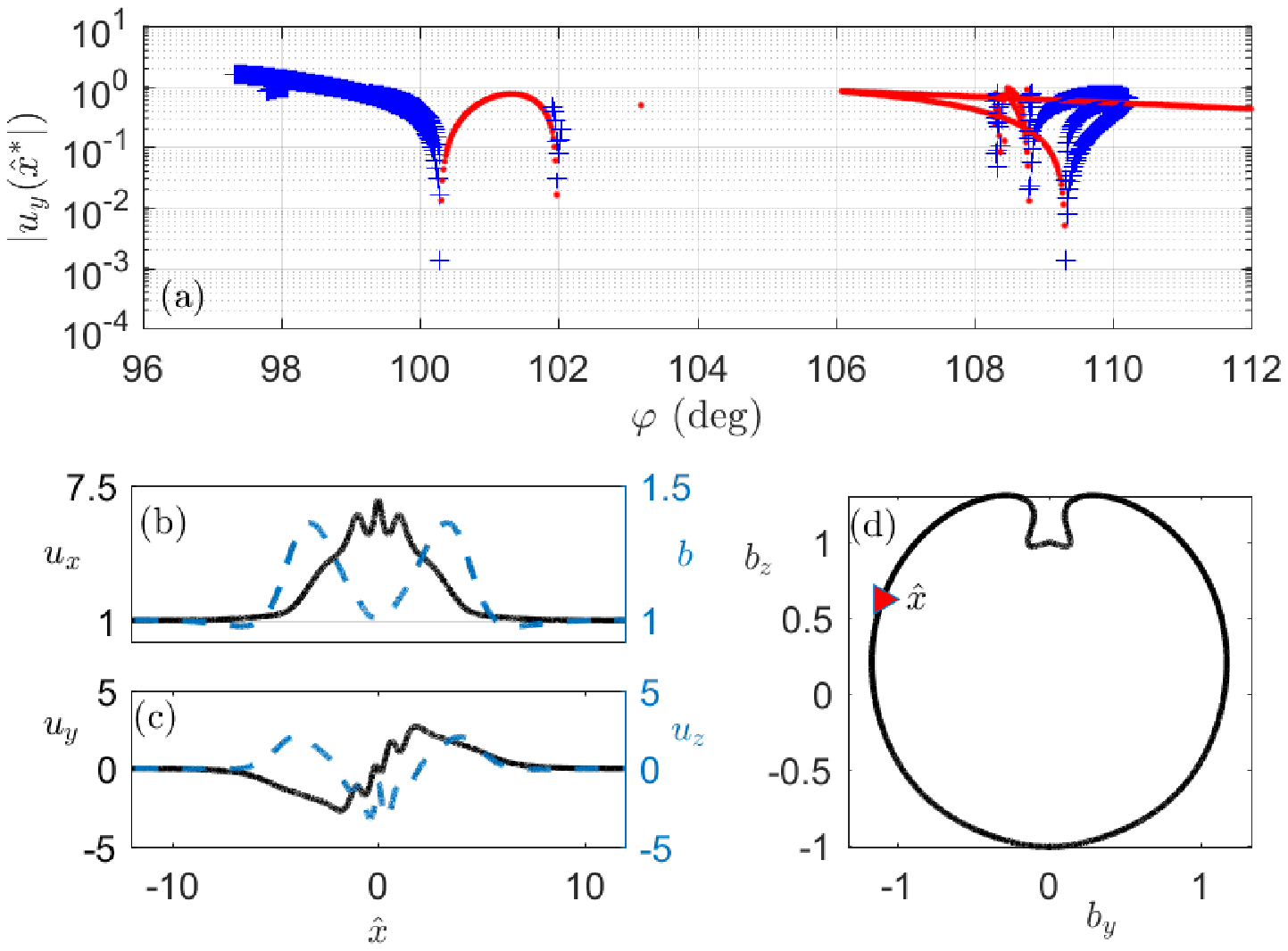}
    \caption{$\mid u_y(\hat{x}^*)\mid$  versus $\varphi$ diagram [panel (a)] and example of solitary wave in the focus-focus domain [panels (b)-(d)].}
    \label{Fig:Focus:Focus}
\end{figure}

In order to illustrate this case, we now present some results for
the parameter values of case (2) in Table \ref{Table:Par},  $\theta
= 70^\circ$ and $C/V_A = 0.52$. We set the numerical parameter
$\epsilon=10^{-9}$, and computed the orbits of Eq.
\eqref{Eq:Dynamical:System} with initial conditions given by  Eq.
\eqref{Eq:IC:Focus:Focus} and $\varphi$ from $96^\circ$ to
$112^\circ$. For each of them, the value of $u_y$ at the
intersection with the symmetric section, i.e. $u_y(\hat{x}^*)$, was
computed and presented in a $\varphi$ versus $\mid
u_y(\hat{x}^*)\mid $ diagram [see panel (a) in Fig.
\ref{Fig:Focus:Focus}]. Similarly to the previous section, blue
crosses and red dots were used to denote  positive and negative
values of $u_y(\hat{x}^*)$ and highlight the changes of signs and
locate the existence of a solitary waves. For instance, a solitary
wave exist for $\varphi \approx 100.3^\circ$ and its structure is
given in panels (b)-(d). Interestingly,  it can be proved that the
existence of one solitary wave for a given value of the physical
parameters implies the existence of infinitely many others if the
system is reversible and the upstream state is a focus-focus
\cite{Harterich_1998}. Such a theoretical result, which was
demonstrated earlier for conservative systems \cite{Devaney_76}, is
a consequence of the spiralling linear dynamics due to the complex
eigenvalue and the additional orbits are like copies of the original
one but with extra oscillations. Several of these extra orbits can
be identified in panel (a).

\section{Stability of solitary waves\label{Sec:Stability}}

Previous sections analyzed the organization of the solitary waves in
the propagation angle-velocity plane and discussed their main
physical features such as polarizations and structure. However, just
the existence of these solutions in the FLR-Hall-MHD model does not
guarantee their physical relevance. The observation of these
localized structures in real scenarios, such as the solar wind, is
also linked to the concepts of excitation and stability. Although a
thorough analysis is well beyond the scope of this work, we now
illustrate with few examples some interesting features observed in
non-stationary FLR-Hall-MHD (Eqs.
\eqref{e:continuity_FLR_non}-\eqref{e:magnetic_induction_FLR_non}) simulations 
initialized with exact solitary waves. \EBedit{Then, the stability of some solutions is also investigated for the case of using dynamical pressure equations with FLR work corrections [see Eqs. \eqref{Eq:P:evol:Parallel} and \eqref{Eq:P:evol:Perpen}]. In this case, the initial solution is a non-exact solitary wave solution of the new system}. As a \ESedit{preliminary} step, we first
analyzed the linear stability, i.e. dispersion relation, of the
background plasma state.

\subsection{Dispersion relation of the FLR-Hall MHD system \label{Sec:Dispersion}}

Since the solitary waves satisfy $\bm{\xi}\rightarrow \bm{\xi}_0$ \TPedit{as}
$\hat{x}\rightarrow \pm \infty$, a necessary condition for their
stability is the linear stability of the background plasma state. We
analyze it by writing the fluid variables as
\begin{align}
\hat{\rho}   =& 1+\hat{\rho}_1 e^{i\left(k\hat{x}-\omega\tau\right)}\\
\bm{u} =& \bm{e}_x+\left(\hat{u}_{x1}\bm{e}_y + \hat{u}_{y1}\bm{e}_y + \hat{u}_{z1}\bm{e}_z\right) e^{i\left(k\hat{x}-\omega\tau\right)}\\
\hat{\bm{B}} =& \frac{\bm{e}_x}{\tan\theta} +
\bm{e}_z+\left(\hat{b}_{y1}\bm{e}_y + \hat{b}_{z1}\bm{e}_z\right)
e^{i\left(k\hat{x}-\omega\tau\right)},
\end{align}
where \GSAedit{$\hat{\rho}=\rho/\rho_0$, $\hat{B} =
B/B_0\sin\theta$}, $k$ and $\omega$ represent the normalized
wavevector and frequency of the small perturbations denoted with
subscript $1$. Substituting these expansions in Eqs.
\eqref{e:continuity_FLR_non}-\eqref{e:magnetic_induction_FLR_non}
and retaining only first order terms \TPedit{yields the} homogeneous linear
system
\begin{equation}
\bar{\bm{D}}\left(\omega,k\right)\GSAedit{\hat{\bm{\eta}}} = 0
\end{equation}
with $\GSAedit{\hat{\bm{\eta}}} = [\hat{\rho}_1\ \ \hat{u}_{x1}\ \
\hat{u}_{y1}\ \ \hat{u}_{z1}\ \ \hat{b}_{y1}\ \ \hat{b}_{z1}]$. For
given values of $M_e$, $M_i$, $\theta$, and $a_p$, the compatibility
condition $\det\left(\bar{\bm{D}}\right) = 0$  gives the dispersion
relation $\omega = \omega(k)$. The background state is unstable if
$\omega$ is imaginary and we can compute the growth rate as $\gamma
= \max\left[\Im\left(\omega\right)\right]$.

\begin{figure}[h]
    \centering
    \noindent\includegraphics[scale=.9]{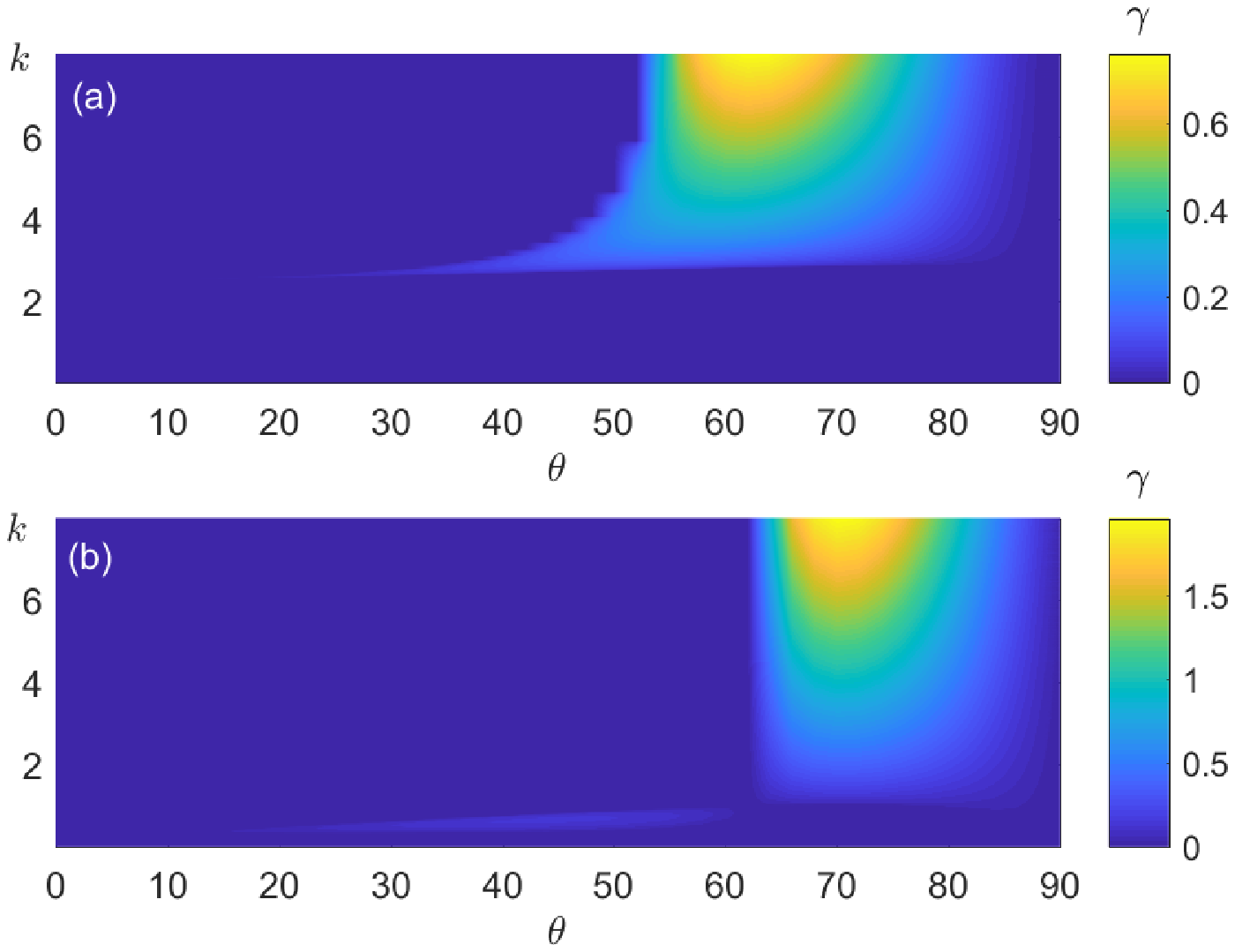}
    \caption{Panels (a) and (b) show the maximum growth rate $\gamma$ in the $k-\theta$ plane for cases 1 and 2 in Table \ref{Table:Par}, respectively.}
    \label{Fig:gamma}
\end{figure}

Panels (a) and (b) in Fig. \ref{Fig:gamma} show  the value of the
growth rate in the $k-\theta$ plane for cases 1 and 2 in Table
\ref{Table:Par}. For this particular set of \GSAedit{parameters} we
conclude that only low-angle propagation waves could be stable. By
comparing this diagram with Fig. \ref{Fig:Branches}, \EBedit{one finds
that the tails of all the waves computed at $\theta > 50^\circ $ are unstable.}
\EBedit{The use of extended precision allowed us to locate solitary wave solutions at lower propagation angles, for which the maximum growth rate of the perturbations to the background state tends to zero.}

\subsection{Numerical Simulations }

\GSAedit{This section studies the stability of two solitary waves
with parameters given by case 1 in Table \ref{Table:Par}. In both
cases, the waves belong to the parameter regime where $\bm{\xi}_0$
is a saddle-center and a relation between $\theta$ and $C/V_A$ holds
(branches of solutions). They were used as initial conditions in
Eqs.
\eqref{e:continuity_FLR_non}-\eqref{e:magnetic_induction_FLR_non}
and their evolutions were found by integrating the equations
numerically with a spectral method, following Ref.
\cite{Laveder_11} (find some
details on the numerical method in Appendix
\ref{Sec:spectral_code}).}

\begin{figure}[h]
	\centering
	\noindent\includegraphics[scale=.9]{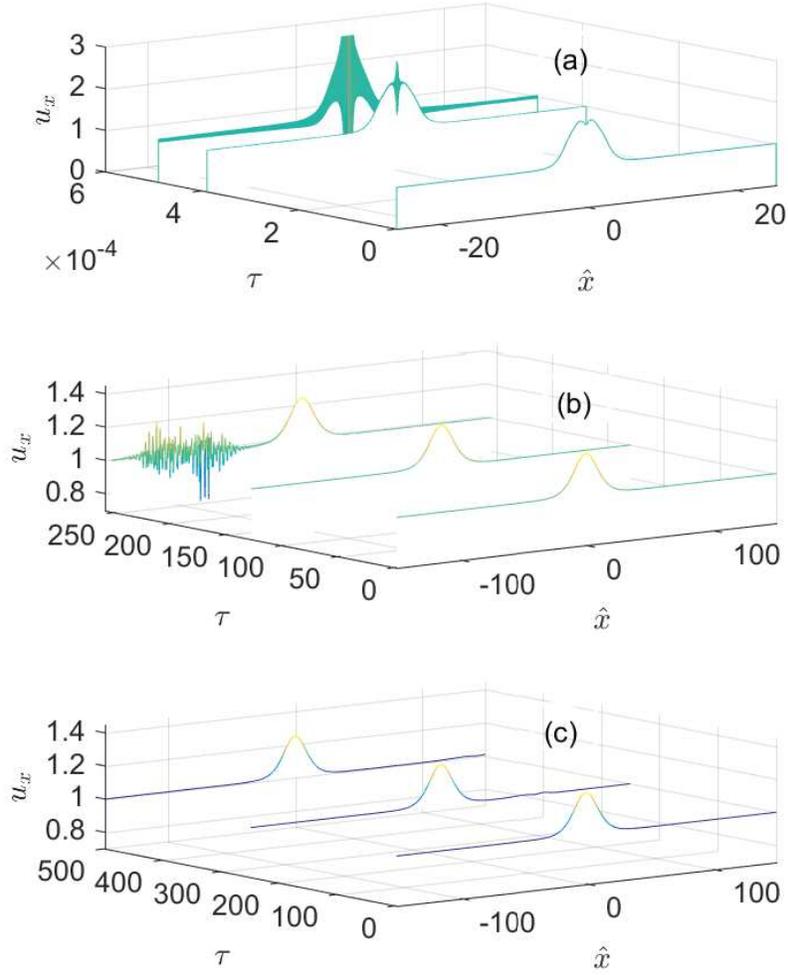}
	\caption{Evolutions of some solitary waves belonging to the saddle-center regime. Panel (a) shows an example of unstable core for $\theta=80^\circ$ and $C/v_A = 0.745715$. Panel (b) shows an example of unstable background for $\theta=30.415^\circ$ and $C/v_A = 0.9$. Panel (c) shows an example of robust solution, using Eqs. \eqref{Eq:P:evol:Parallel} and \eqref{Eq:P:evol:Perpen} for the pressures with the same parameters as (b).}
	\label{Fig:Simulations}
\end{figure}

The evolution of the spatial profile of $u_x$ for the first wave,
which has $\theta=80^\circ$ and $C/V_A = 0.745715$, is shown in
panel (a) of Fig. \ref{Fig:Simulations}. According to Fig.
\ref{Fig:gamma} the background plasma state is unstable for such a
high propagation angle. However, as shown in Fig.
\ref{Fig:Simulations}, the core of the solitary wave is unstable and
the solitary wave is destroyed even before the instability at the
tail would be developed. The behavior of the second wave,  having \EBedit{velocity $C/V_A = 0.9$ and a
propagation angle $\theta = 30.415^\circ$}, is totally different [see
panel (b)]. For this case, the core of the solitary wave is stable
and the instability happens at the tail. The results of the
simulation, i.e. the most unstable wavevector and the growth rate,
are consistent with the analysis of Sec. \ref{Sec:Dispersion}.
Interestingly, although the wave is unstable, the core of this solitary
wave is quite robust and survives a time longer than
\EBedit{$250M_A\cos\theta/\Omega_{ci,0}\approx 266
\Omega_{ci,0}^{-1}$}.

The simulation results for the second wave show that the instability
may come from the unstable character of the background plasma state
in the framework of the FLR-Hall-MHD system closed with a double
adiabatic pressure model. For this reason, we investigated a bit
further the stability of the second wave by repeating the
simulations but now using the dynamic equations for the pressures.
\GSAedit{Following Ref. \cite{Sulem_15} (find a short discussion in
Appendix \ref{Sec:App:Dynamic:Pressure}), we write}
\begin{align}
\nonumber\frac{\partial P_{\parallel}}{\partial \tau} &+ \hat{\nabla} \cdot \left(P_{\parallel} \mathbf{u}\right) + 2 P_{\parallel} \mathbf{e}_b \cdot \hat{\nabla} \mathbf{u} \cdot \mathbf{e}_b + \\ &+ \frac{1}{a_p} \left[ \left( \bar{\mathbf{\Pi}} \cdot \hat{\nabla} \mathbf{u} \right)^S : \bar{\bm{\tau}} -  \bar{\mathbf{\Pi}} : \frac{d\bar{\bm{\tau}}}{d\tau} \right] = 0 \label{Eq:P:evol:Parallel}\\
\nonumber\frac{\partial P_{\perp}}{\partial \tau} &+ \hat{\nabla} \cdot \left(P_{\perp} \mathbf{u}\right) + P_{\perp} \hat{\nabla} \cdot \mathbf{u} -  P_{\perp} \mathbf{e}_b \cdot \hat{\nabla} \mathbf{u} \cdot \mathbf{e}_b + \\ \nonumber & + \frac{1}{2} \left[ \left( \bar{\mathbf{\Pi}} \cdot \hat{\nabla} \mathbf{u} \right)^S : \bar{\mathbf{I}} - \left( \bar{\mathbf{\Pi}} \cdot \hat{\nabla} \mathbf{u} \right)^S : \bar{\bm{\tau}}\right. + \\ &\left. +  \bar{\mathbf{\Pi}} : \frac{d\bar{\bm{\tau}}}{d\tau}\right]  = 0 \label{Eq:P:evol:Perpen}
\end{align}
where $\hat{\nabla} = \partial/\partial \hat{x}$,  $\bar{\bm{\tau}}
= \mathbf{e}_b \mathbf{e}_b$ and $\bar{\bm{\Pi}}$ is the
nondimensional ion FLR pressure tensor, given by Eqs.
\eqref{e:Pi11}-\eqref{e:Pi31} scaled with $p_{\perp0}$.  The
superscript $S$ means that  the tensor  between parentheses is
symetrized by the addition of its transpose. Unlike
the double adiabatic model, Sys.
\eqref{e:continuity_FLR_non}-\eqref{e:magnetic_induction_FLR_non}
and Eqs. \eqref{Eq:P:evol:Parallel}-\eqref{Eq:P:evol:Perpen}
conserves the energy
\begin{align}
\nonumber E &= \int_{-\infty}^{\infty} \left[  \frac{1}{2}\hat{\rho} \mathbf{u}^2 + \frac{1}{2} M_A \sin^2 \theta b^2 + M_e \hat{\rho} \log \hat{\rho} + \right. \\ &\left.+ M_i \left( P_{\perp} + \frac{1}{2} a_p P_{\parallel} \right) \right] d\hat{x}
\end{align}

As shown in \ESedit{Fig.~\ref{Fig:Simulations}}, panel (c), the substitution of the crude double
adiabatic approximation by the pressure evolution equation
suppresses the instability of the wave. Since the initial condition
is not an exact solution of the FLR-Hall-MHD model closed with Eqs.
\eqref{Eq:P:evol:Parallel}-\eqref{Eq:P:evol:Perpen}, the wave is
distorted slightly but it still propagates for \EBedit{times longer than 500 $M_A\cos\theta/\Omega_{ci,0}$} while keeping its original shape. \EBedit{The time integration was stopped at 500 $M_A\cos\theta/\Omega_{ci,0}$ but the simulation was still stable. The shape of the wave at this time is practically identical to the given initial condition and it propagates with the speed of the used reference frame (the wave does not drift). These results suggest that this particular solitary wave computed with the double adiabatic pressure model is very close to be an exact solution of the system with dynamical pressure equations. Similar simulations performed at $\theta=50^\circ$ showed less robust behavior, greater level of deformation and drift leftwards of the domain.} 

\section{Conclusions\label{Sec:Conclusions}}

The existence of low frequency solitary waves in magnetized plasmas
is firmly supported by space observations. For this reason, the
knowledge of the physical properties of these structures, including
amplitudes, spatial profiles of the fluid and electromagnetic
fields, and polarizations, are relevant. The existence of possible
relations linking physical parameters, such as propagation angle of
the wave with respect to the ambient magnetic field $\theta$ and the
propagation velocity $C$, are also important because they can be
helpful interpreting the experimental data. The answers to most of
these interesting questions can be obtained by analyzing the
dynamical system obtained from the double adiabatic FLR-Hall-MHD
model after assuming the 1-dimensional traveling wave ansatz.

First, solitary waves can exist if the background plasma state,
which appears in the dynamical system as an equilibrium state, is
not a center-center. Moreover, using simple geometrical arguments
based on the effective dimension of the dynamical system and its
reversible character, the organization of the waves in the
$\theta-C$ plane can be anticipated even before computing them. If
the background plasma state is a saddle-saddle or a focus-focus the
spectrum of the waves is continuous and, in case it is a
saddle-center, they are organized in branches (relations of the type
$C=C(\theta)$). The numerical scheme (bisection method) presented in
this work can be used to find solitary waves in any of these regions
and proves their existence rigorously. Abundant solitary waves,
including dark and bright waves and some of them belonging to
regions where they were not found before, were computed. The
structures of seven waves were presented in detail and some
differences with respect to the Hall-MHD case (without FLR effects)
were highlighted. Nevertheless, deeper parametric analysis are
necessary to construct a more complete picture about the properties
and organization of the waves in parameters space. For instance, the
fact that we did not find waves with banana-like polarization, a
very peculiar signature observed in the solar wind and in more
simple theoretical models, does not preclude their existence in this
FLR-Hall-MHD model. Another topic that could  be investigated in
future works is the analysis of the existence of the so-called
quasi-solitons, i.e. a more general class of solutions that would
contain the branches of solutions found in this work as a particular
subclass.

Regarding the stability of the waves, a linear analysis within the
framework of the double adiabatic FLR-Hall-MHD model shows that the
background plasma state is unstable for the parameters under
consideration. Since the tails of the waves approach to such state
at plus and minus infinity, they are also unstable. However, some
numerical simulations indicate that the core can be  stable.
Moreover, the substitution of the double adiabatic model by
evolution equations for the pressures shows that some solitary
waves can be robust. This result opens the interesting problem about
the computation of exact solitary waves in the framework of the
FLR-Hall-MHD with evolution equations for the pressures. Such a
study, which is beyond the scope of the present work, is challenging
because the dimension of the phase space of the dynamical system
would be larger and several of the geometrical arguments used in
this work should be revised.

\begin{acknowledgments}
G.S.A. is supported by the Ministerio de Econom\'ia y Competitividad
of Spain under the Grant No RYC-2014-15357. ES was supported by the 
Swedish Research Council, Grant No. 2016-05012 and by the Knut and Allice Wallenberg Foundation.
\end{acknowledgments}

\appendix

\section{Dynamical System\label{Sec:Dynamical:System}}

This section follows Ref. \cite{Mjolhus_2009} to find the explicit
form of vector $\mathbf{f}$ in Eq. \eqref{Eq:Dynamical:System}. For
convenience, we split such a column vector as $\mathbf{f} =
[\mathbf{f}_u \ \ \mathbf{f}_b] $, with $\mathbf{f}_u$ and
$\mathbf{f}_b$ the three and two-dimensional vector flows governing
the dynamics of $\bm{u}$ and $\bm{b}$, respectively. Equations
\eqref{e:continuity_FLR}-\eqref{e:magnetic_induction_FLR} are
particularize to one-dimensional ($\partial/\partial y =
\partial/\partial y = 0$) and steady ($\partial/\partial t = 0$)
solutions. Equation \eqref{e:continuity_FLR} then becomes
\begin{equation}
\label{e:rho} \frac{\rho}{\rho_0} = \frac{1}{u_x}
\end{equation}
After defining the new variables $P_{\parallel} (u, b^2)\equiv
p_\perp/p_{\perp0}$ and $P_{\perp} (u, b^2) \equiv
p_\parallel/p_{\parallel 0}$, the equations of state
\eqref{e:p_par_Hall} and \eqref{e:p_perp_Hall} read
\begin{equation}
P_{\parallel} (u_x, b^2) = \frac{1}{\hat{b}^2 u_x^3}, \ \ \ \ \ \ \
P_{\perp} (u_x, b^2)= \frac{\hat{b}}{u_x} \label{Eq:double_ad}
\end{equation} where we introduced the dimensionless quantities $b^2
= b_y^2 + b_z^2$ and $\hat{b}^2 = \left(B/B_0\right)^2 = \cos^2
\theta + b^2 \sin^2 \theta$.

The component of Eq. \eqref{e:magnetic_induction_FLR} along the
propagation direction $x$ gives $B_x = B_0\cos\theta$. In the
transverse direction  one finds
\begin{equation}
\frac{d}{d \hat{x}} \left[ u_x b_z \sin \theta  - u_z \cos \theta  +
u_x \sin \theta  \frac{d b_y }{d \hat{x}} \right] = 0
\end{equation}
\begin{equation}
\frac{d}{d \hat{x}} \left[ u_x b_y \sin \theta  - u_y \cos \theta  -
u_x \sin \theta  \frac{d b_z}{d \hat{x}} \right] = 0
\end{equation}
Using the the plasma conditions upstream, this set of equations are
integrated to find the two-dimensional flow
\begin{equation}
 \bm{f}_b =\left(\begin{array}{c} \frac{u_z}{u_x} \frac{\cos \theta}{\sin
\theta} - b_z + \frac{1}{u_x}\\
-\frac{u_y}{u_x} \hspace{1mm}\frac{\cos \theta}{\sin \theta} + b_y
\end{array}\right)\label{e:Gy}
\end{equation}
This flow coincides with Eq. (17) in Ref. \cite{Mjolhus_2009},
except for the term $1/u_x$ in the first row of Eq. \eqref{e:Gy}.

Following a similar procedure, Eq. \eqref{e:momentum_FLR} gives ,
\begin{equation}
\label{e:du_vec_dx}  \bar{\mathbf{A}} \cdot \frac{d \mathbf{u}}{d
\hat{x}} + \mathbf{F} (\mathbf{u}, \mathbf{b}) = 0
\end{equation}\\
where we introduced the flow $\mathbf{F} =\left( F_x \mathbf{e}_x +
F_y \mathbf{e}_y + F_z \mathbf{e}_z\right)/\delta$ with
\begin{align}
 F_x   =& u_x - 1 + P(u_x, b^2) + \frac{1}{2} M_A \sin^2 \theta \left( b^2 - 1 \right) \label{e:Fx}\\
 F_y   =& u_y + \chi (u_x, b^2) \cos \theta \sin \theta \hspace{1mm} b_y\label{e:Fy}\\
 F_z   =& u_z + \left[ \chi(u_x, b^2) b_z - \chi(1,1) \right] \cos \theta \sin \theta\label{e:Fz}
\end{align}
and the auxiliary functions
\begin{align}
\delta =& \frac{M_i}{M_A} \frac{P_{\perp}\left(u_x,b^2\right)}  {\hat{b} \cos \theta}\label{Eq:delta}\\
P(u_x, b^2) =&\ M_e \left( \frac{1}{u_x} - 1 \right) + M_i \left\lbrace P_{\perp} (u_x,b^2) - 1 + \vphantom{\frac12} \right. \\
        &  + \left[ a_p P_{\parallel} (u_x, b^2) - P_{\perp} (u_x,b^2) \right] \frac{\cos^2 \theta}{\hat{b}^2}
        \nonumber \\
        & \left. - \left(a_p - 1\right) \cos^2 \theta \right\rbrace
        \label{e:P}\\
\chi(u_x, b^2) =&\ M_i \left[ a_p P_{\parallel} (u,b^2) - P_{\perp}
(u,b^2) \right] \frac{1}{\hat{b}^2} - M_A  \label{e:chi}
\end{align}
Factor $1/\hat{b}$ appearing in Eq. \eqref{Eq:delta}, which comes
from the fact that $\Omega_{ci}$ in Eqs.
\eqref{e:Pi11}-\eqref{e:Pi31} is the local ion gyro frequency, is
missed in Ref. \cite{Mjolhus_2009}. Tensor $\bar{\mathbf{A}}$ in
Equation \eqref{e:du_vec_dx} is
\begin{equation}
\bar{\mathbf{A}} = \bar{\mathbf{I}} \times \mathbf{r} - 2\varepsilon
\hat{\mathbf{b}} \hat{\mathbf{b}}_{\parallel} \times
\hat{\mathbf{b}}_{\perp}\label{Eq:Tensor:A}
\end{equation}
with $\hat{\mathbf{b}}_{\parallel} = \hat{b}_{x} \mathbf{e}_x $,
$\hat{\mathbf{b}}_{\perp} = \hat{b}_y \mathbf{e}_y + \hat{b}_z
\mathbf{e}_z$, $\hat{\mathbf{b}} = \hat{\mathbf{b}}_{\parallel} +
\hat{\mathbf{b}}_{\perp}$, $\hat{b}_x =  B_x / B = \cos \theta /
\hat{b}$ and $\hat{b}_{y,z} = B_{y,z}/B = b_{y,z} \sin \theta /
\hat{b}$, and
\begin{flalign}
&\label{e:r_vec} \mathbf{r} = -r_{\parallel} \hat{\mathbf{b}}_{\parallel} + r_{\perp} \hat{\mathbf{b}}_{\perp} \\
&\label{e:r_par} r_{\parallel} = \frac{1}{2} \left(1 - 3 \hat{b}_{\parallel}^2\right) + 2 \varepsilon \hat{b}_{\parallel}^2 \\
&\label{e:r_perp} r_{\perp} = \frac{1}{2} \left(1 + 3
\hat{b}_{\parallel}^2\right) - 2 \varepsilon \hat{b}_{\parallel}^2\\
&\label{e:epsilon} \varepsilon = \left(p_{\perp} -
p_{\parallel}\right) / p_{\perp} = 1 - a_p P_{\parallel}/P_{\perp}
\end{flalign}
Following Ref. \cite{Mjolhus_2009}, tensor $\bar{\mathbf{A}}$ will
be referred as the 1-FLR tensor. We mention that a plus sign
(instead a minus) was written in the second term of Eq.
\eqref{Eq:Tensor:A} in Ref. \cite{Mjolhus_2009}.

%    \textcolor{red}{Tensor $\bar{\mathbf{A}}$ por componentes, en caso de que fuese útil:
%    \begin{flalign*}
%    &A_{11} = 0 \\
%    &A_{21} = \hat{b}_z r_{\perp} \\
%    &A_{31} = -\hat{b}_y r_{\perp} \\
%    &A_{12} = -\hat{b}_z \left( r_{\perp} - 2 \varepsilon \hat{b}_{\parallel}^2 \right)\\
%    &A_{22} = 2\varepsilon \hat{b}_{\parallel} \hat{b}_{y} \hat{b}_{z}\\
%    &A_{32} = -\hat{b}_{\parallel} \left( r_{\parallel} - 2\varepsilon \hat{b}_z^2 \right)\\
%    &A_{13} = \hat{b}_y \left( r_{\perp} - 2 \varepsilon \hat{b}_{\parallel}^2 \right)\\
%    &A_{23} = \hat{b}_{\parallel} \left( r_{\parallel} - 2\varepsilon \hat{b}_y^2 \right)\\
%    &A_{33} = -2\varepsilon \hat{b}_{\parallel} \hat{b}_y \hat{b}_z
%    \end{flalign*}
%    }

\subsection{Singularity of the tensor $\bar{\mathbf{A}}$ and invariant \GSAedit{manifold} \label{Sec:Singularity}}

As pointed out in Refs. \cite{Hau_1991} and \cite{Mjolhus_2009},
tensor $\bar{\mathbf{A}}$ is singular
\begin{align}
 \mathbf{L} \cdot \bar{\mathbf{A}} = 0 \\
\bar{\mathbf{A}} \cdot \mathbf{R} = 0
\end{align}
and left and right null vectors are given by
\begin{align}
\mathbf{L} =& \frac{1}{\mu} \left( \mathbf{r} + 2 \varepsilon \hat{b}_{\perp}^2 \hat{\mathbf{b}}_{\parallel} - 2 \varepsilon \hat{b}_{\parallel}^2 \hat{\mathbf{b}}_{\perp} \right) \label{e:L}\\
\mathbf{R} =& \frac{1}{\mu} \mathbf{r}.  \label{e:R}
\end{align}
After imposing the condition $\mathbf{L} \cdot \mathbf{R} = 1$, the
arbitrary constant $\mu$ is
\begin{align} \mu^2 = r_{\perp}^2
\hat{b}_{\perp}^2 + r_{\parallel}^2 \hat{b}_{\parallel}^2 - \gamma,
\label{e:mu}
\end{align}
with $\gamma = 2 \varepsilon \hat{b}_{\perp}^2
\hat{b}_{\parallel}^2$. A direct result of the singular character of
$\bar{\mathbf{A}}$ is the constraint
\begin{equation}
H (\bm{\xi}) = \mathbf{L} \cdot \mathbf{F} = 0, \label{e:invariant}
\end{equation}
which is easily obtained by dotting Eq. \eqref{e:du_vec_dx} from the
left with $ \mathbf{L}$. Therefore, any orbit of the
five-dimensional phase space of $\bm{\xi}$ in Eq.
\eqref{Eq:Dynamical:System} should lie in the four-dimensional
surface defined by the constraint $H$. As a consequence, the
effective dimension of the system is four. Although a detailed
derivation on how Eq. \eqref{Eq:Dynamical:System} can be obtained
from \eqref{e:du_vec_dx} was given in Ref. \cite{Mjolhus_2009}, we
summarize below the most important calculations because some small
discrepancies were found.

Besides the zero eigenvalue, the 1D FLR tensor $\bar{\mathbf{A}}$
has imaginary eigenvalues $\pm i \mu$, where $\mu$ is given by Eq.
\eqref{e:mu}. It can be shown \cite{Mjolhus_2009} that vectors
\begin{align}
\mathbf{S} =& \frac{1}{\mu \hat{b}_{\perp} \hat{b}_{\parallel}}
\left[ \left( \hat{b}_{\perp}^2 r_{\perp} - \gamma \right)
\hat{\mathbf{b}}_{\parallel} +  \left( \hat{b}_{\parallel}^2
r_{\parallel} - \gamma \right) \hat{\mathbf{b}}_{\perp} \right] \label{e:S} \\
\mathbf{T} =& \frac{1}{\hat{b}_{\perp} \hat{b}_{\parallel}} \left(
\hat{\mathbf{b}}_{\parallel} \times \hat{\mathbf{b}}_{\perp}
\right)\label{e:T} \\
\mathbf{M} =& \frac{1}{\mu \hat{b}_{\perp} \hat{b}_{\parallel}} \left( \hat{b}_{\perp}^2 r_{\perp} \hat{\mathbf{b}}_{\parallel} + \hat{b}_{\parallel}^2 r_{\parallel} \hat{\mathbf{b}}_{\perp} \right)\label{e:M} \\
 \mathbf{N} =&  \mathbf{T} \label{e:N}
    \end{align}
satisfy the relations $\bar{\mathbf{A}} \cdot \mathbf{S} = -\mu
\mathbf{T}$, $\bar{\mathbf{A}} \cdot \mathbf{T} = \mu \mathbf{S}$, $
\mathbf{M} \cdot \bar{\mathbf{A}} = \mu \mathbf{N} $ and $\mathbf{N}
\cdot \bar{\mathbf{A}} = -\mu \mathbf{M}$. One also readily finds
that the following orthogonality and normalization conditions hold
\begin{align}
&\mathbf{L} \cdot \mathbf{R} = \mathbf{M} \cdot \mathbf{S} = \mathbf{T} \cdot \mathbf{N} = 1\\
&\mathbf{M} \cdot \mathbf{R} = \mathbf{N} \cdot \mathbf{R} =
\mathbf{L} \cdot \mathbf{S} = \mathbf{L} \cdot \mathbf{T} =
\mathbf{N} \cdot \mathbf{S} = \mathbf{M} \cdot \mathbf{T} = 0
\end{align}

The new base $\mathbf{R}$, $\mathbf{S}$, $\mathbf{T}$ will allow us
to find the flow $\mathbf{f}_u$ in Eq. \eqref{Eq:Dynamical:System}
from Eq. \eqref{e:du_vec_dx}. We first decompose $\mathbf{f}_u$ and
$\mathbf{F}$ on that base and write
\begin{align}
\mathbf{F} = F_R \hspace{1mm} \mathbf{R} + F_S \hspace{1mm} \mathbf{S} + F_T  \hspace{1mm} \mathbf{T}\label{e:F_RST} \\
\mathbf{f}_u = f_{uR}  \mathbf{R} + f_{uS} \mathbf{S} + f_{uT}
\mathbf{T}\label{e:w_RST}
\end{align}
The dot product of Eq. \eqref{e:F_RST} by $\mathbf{L}$, $\mathbf{M}$
and $\mathbf{N}$  gives
    \begin{flalign}
    & \label{e:FR} \hspace{1cm} F_R = \mathbf{L} \cdot \mathbf{F} = 0 \\
    & \label{e:FS} \hspace{1cm} F_S = \mathbf{M} \cdot \mathbf{F} \\
    & \label{e:FT} \hspace{1cm} F_T = \mathbf{N} \cdot \mathbf{F}
    \end{flalign}
where we used Eq. \eqref{e:invariant} and the orthogonality and
normalization conditions. We now find the components of
$\mathbf{f}_u$ by first noting that Eqs. \eqref{Eq:Dynamical:System}
and \eqref{e:du_vec_dx} give
\begin{equation}
\bar{\mathbf{A}} \cdot\mathbf{f}_u = -\mathbf{F}\label{Eq:A:fu:F}
\end{equation}
The substitution in Eq. \eqref{Eq:A:fu:F} of Eq. \eqref{e:w_RST} and
the use of Eqs. \eqref{e:S} and \eqref{e:T} yield
\begin{align}
f_{uS} = F_T / \mu
\label{e:wS}\\
f_{uT} = - F_S / \mu . \label{e:wT}
\end{align}
Finally, the component $f_{uR}$ is found from the constraint
\eqref{e:invariant}. From such invariant, one finds
\begin{equation}
\frac{dH}{d \hat{x}} = \frac{\partial H}{\partial \mathbf{u}} \cdot
\frac{d\mathbf{u}}{d \hat{x}} + \frac{\partial H}{\partial
\mathbf{b}} \cdot \frac{d\mathbf{b}}{d \hat{x}} =  \frac{\partial
H}{\partial \mathbf{u}} \cdot \mathbf{f}_u + \frac{\partial
H}{\partial \mathbf{b}} \cdot \mathbf{f}_b = 0\label{e:dH_dx}
\end{equation}
where we used Eq. \eqref{Eq:Dynamical:System}. The component
$f_{uR}$ then reads
\begin{equation}
f_{uR} = -\frac{\Gamma_S f_{uS}+\Gamma_b}{\Gamma_R}\label{Eq:fur}
\end{equation}
with
\begin{equation}
\Gamma_R = \frac{\partial H}{\partial \mathbf{u}} \cdot \mathbf{R},
\ \ \ \ \   \Gamma_S = \frac{\partial H}{\partial \mathbf{u}} \cdot
\mathbf{S}, \ \ \ \ \
  \Gamma_b = \frac{\partial H}{\partial \mathbf{b}} \cdot
  \mathbf{f}_b\label{Eq:Gamma}
\end{equation}
and \TPedit{where} we used that $\partial H/\partial\bm{u}\cdot \bm{T}=0$ because
(i) $\mathbf{L} \cdot \mathbf{T} = 0$, (ii) the derivatives of
$\mathbf{L}$ with respect to $u_y$ and $u_z$ are all zero and
$\partial \mathbf{L}/ \partial \mathbf{u} \cdot \mathbf{F}$ is along
$\mathbf{e}_x$, and (iii) as shown by Eq. \eqref{e:T}, $\mathbf{T}$
is perpendicular to $\mathbf{e}_x$. The analytical derivatives of
$\frac{\partial H}{\partial \mathbf{u}}$ and $\frac{\partial
H}{\partial \mathbf{b}}$ have been implemented in our code.
Equations \eqref{e:dH_dx} and \eqref{Eq:fur} have a sign different
as compared with the corresponding equations in Ref.
\cite{Mjolhus_2009}.

The initial conditions used in this work are consistent with the
constraint \eqref{e:invariant} because $H(\bm{\xi}_0)=0$.  The flow
$\bm{f}$ in Eq. \eqref{Eq:Dynamical:System} guarantees that the
orbit $\bm{\xi}(\hat{x})$ will lie in the manifold $H=0$. As pointed
out in Ref. \cite{Mjolhus_2009}, orbits cannot cross the set $U$
defined by $\Gamma_R(U) = 0$, which plays a similar role to the
sonic circle in the Hall-MHD theory \cite{Mjolhus_06}.

\section{1-Dimensional FLR-Hall MHD spectral code}
\label{Sec:spectral_code}

For convenience, the simulations in Sec. \ref{Sec:Stability}   used
the same dimensionless variables as  in previous section and also
$\hat{\rho}=\rho/\rho_0$ and  the normalized time $\tau =
v_{x0}t/\ell$. After substituting $\nabla =
\partial /
\partial x \hspace{1mm}\mathbf{e}_x$, Eqs.
\eqref{e:continuity_FLR}-\eqref{e:magnetic_induction_FLR} becomes
\begin{equation}
\frac{\partial \hat{\rho}}{\partial \tau} + \frac{\partial}{\partial
\hat{x}} \left(\hat{\rho} u_x \right) = 0
\label{e:continuity_FLR_non}
\end{equation}
\begin{align}
 \frac{\partial}{\partial
\tau} \left( \hat{\rho} \mathbf{u} \right) + \bm{e}_x\cdot
\frac{\partial}{\partial\hat{x}} \left[ \hat{\rho} \mathbf{u}
\mathbf{u} + M_e \hat{\rho} \bar{\mathbf{I}} + M_i
\hat{\bar{\mathbf{P}}}_i^{\left(0\right)} \right. + \nonumber \\
\left. + M_A \sin^2 \theta \left( \frac{1}{2} \hat{B}^2
\bar{\mathbf{I}} - \hat{\mathbf{B}} \hat{\mathbf{B}}\right) +
M_i \bar{\bm{\Pi}} \right] = 0
\label{e:momentum_FLR_non}
\end{align}
\begin{equation}
 \frac{\partial
\hat{\mathbf{B}}}{\partial \tau} =
\bm{e}_x\times\left[\frac{\partial}{\partial\hat{x}}\left(
\mathbf{u} \times \hat{\mathbf{B}} - \frac{1}{\hat{\rho}}
\frac{\partial \mathbf{b}}{\partial \hat{x}}
\right)\right]\label{e:magnetic_induction_FLR_non}
\end{equation}
with
\begin{align}
\hat{\bar{\mathbf{P}}}_i^{\left(0\right)} =& \frac{a_p \hat{\rho}^3}{\hat{b}^2} \mathbf{e}_b \mathbf{e}_b + \hat{\rho} \hat{b} \left( \bar{\mathbf{I}} - \mathbf{e}_b \mathbf{e}_b \right)\label{e:pressures_disp} \\
 \hat{\mathbf{B}} =& \frac{\mathbf{e}_x}{\tan \theta}  + \mathbf{b}
\end{align}
Tensor $\bar{\bm{\Pi}}$ accounts for the FLR effect and only its
first row
\begin{equation}
\mathbf{e}_x \cdot \bar{\bm{\Pi}} = \frac{1}{M_A \cos \theta} \frac{1}{\hat{b}} \bar{\mathbf{M}}
\cdot \frac{\partial \mathbf{u}}{\partial \hat{x}} \label{e:Pi_ex}
\end{equation}\\
is needed, where $\bar{\mathbf{M}} = P_{\perp} \bar{\mathbf{A}}$.
\GSAedit{After using the double adiabatic equations}, tensor
$\bar{\mathbf{A}}$ is given by Eq. \eqref{Eq:Tensor:A} but with
$\epsilon$ \GSAedit{now taking the form}  $\varepsilon =
1-a_p\hat{\rho}^2/\hat{b}^3$. The components of $\bar{\mathbf{M}}$
read
\begin{align}
&M_{11} = 0 \label{Eq:M:first}\\
&M_{12} = -\frac{1}{2} \hat{b}_z \left[ P_{\perp} + \left(8 a_p P_\parallel - 5 P_{\perp}\right) \hat{b}^2_x \right] \\[3mm]
&M_{13} = \frac{1}{2} \hat{b}_y \left[ P_{\perp} + \left(8 a_p P_\parallel - 5 P_{\perp}\right) \hat{b}^2_x \right]\\[3mm]
&M_{21} = \frac{1}{2} \hat{b}_z \left[ P_{\perp} + \left(4 a_p P_\parallel -  P_{\perp}\right) \hat{b}^2_x \right]\\[3mm]
&M_{22} = 2 \hat{b}_x \hat{b}_y \hat{b}_z \left( P_\perp - a_p P_\parallel \right)\\[3mm]
&\nonumber M_{23} = \frac{1}{4} \hat{b}_x \left[ P_\perp \left( 3 + \hat{b}_x^2 - 9 \hat{b}_y^2 - \hat{b}_z^2 \right) \right. + \\ &\qquad\left. + 8 a_p P_{\parallel} \left(\hat{b}^2_y - \hat{b}^2_x\right) \right] \\[3mm]
&M_{31} = \frac{1}{2} \hat{b}_y \left[ P_{\perp} \left( \hat{b}^2_x - 1 \right) - 4 a_p P_\parallel \hat{b}^2_x \right] \\[3mm]
&\nonumber M_{32} = -\frac{1}{4} \hat{b}_x \left[ P_\perp \left( 3 + \hat{b}_x^2 -  \hat{b}_y^2 - 9 \hat{b}_z^2 \right) + \right. \\ &\qquad \left. + 8 a_p P_{\parallel} \left(\hat{b}^2_z - \hat{b}^2_x\right) \right]\\[3mm]
&M_{33} = -2 \hat{b}_x \hat{b}_y \hat{b}_z \left( P_\perp - a_p P_\parallel \right) \label{Eq:M:last}
\end{align}
Equations \eqref{Eq:M:first}-\eqref{Eq:M:last} have been written in
terms of $P_\perp$ and $P_\parallel$. This is convenient since
several closures for the ion pressure are being used at different
stages of the work.

Equations
\eqref{e:continuity_FLR_non}-\eqref{e:magnetic_induction_FLR_non}
have been integrated numerically with the spectral method (see e.g.
\cite{Laveder_11}). The size of the simulation box and the number of
points of the mesh (after desaliasing by a factor two) were equal to
$94.328$ and 2048 respectively. A spectral cutoff is imposed on
the spectrum at half the spectral domain.

\section{FLR work in dynamical pressure equations\label{Sec:App:Dynamic:Pressure}}

\GSAedit{This section provides explicit equations for the terms
appearing in the right hand side of Eqs. \eqref{Eq:P:evol:Parallel}
and \eqref{Eq:P:evol:Perpen}. Particularizing for $\hat{\nabla} =
\partial/\partial \hat{x}$ and after some development, the FLR work
terms in these equations read}
\begin{align}
\left( \bar{\mathbf{\Pi}} \cdot \hat{\nabla} \mathbf{u} \right)^S : \bar{\bm{\tau}} &= 2 \left(\mathbf{e}_b \cdot \bar{\bm{\Pi}} \cdot \mathbf{e}_x \right) \left( \mathbf{e}_b \cdot \frac{\partial \mathbf{u}}{\partial \hat{x}} \right) \label{Eq:FLR_work:1} \\
\left( \bar{\mathbf{\Pi}} \cdot \hat{\nabla} \mathbf{u} \right)^S : \bar{\mathbf{I}} &= 2 \frac{\partial \mathbf{u}}{\partial \hat{x}} \cdot \bar{\bm{\Pi}} \cdot \mathbf{e}_x \label{Eq:FLR_work:2}\\
\bar{\bm{\Pi}} : \frac{d\bar{\bm{\tau}}}{d \tau} &= 2 \mathbf{e}_b \cdot \bar{\bm{\Pi}} \cdot \mathbf{V}, \label{Eq:FLR_work:3}
\end{align}
\GSAedit{where we introduced the vector}
\begin{equation} \mathbf{V} =
\hat{b}_x \frac{\partial\mathbf{u}}{\partial \hat{x}} -
\frac{1}{\hat{B}} \mathbf{e}_x \times  \frac{\partial
\hat{\mathbf{E}}_H}{\partial \hat{x}} \label{Eq:vector_V}
\end{equation}
\GSAedit{and where $\hat{\mathbf{E}}_H$ is the electric field from the
Hall and electron pressure contributions. It has been normalized
with $c/(v_{0x} B_0 \sin \theta)$ and it takes the form}
\begin{align}
\nonumber\hat{\mathbf{E}}_H &= \frac{\tan \theta}{\hat{\rho}} \left( \mathbf{e}_x \times \frac{\partial \hat{\mathbf{B}}}{\partial \hat{x}} \right)  \times \hat{\mathbf{B}} + \\ &+ \frac{M_e}{M_A \sin \theta \cos \theta} \frac{1}{\hat{\rho}} \frac{\partial \hat{\rho}}{\partial \hat{x}} \mathbf{e}_x .\label{Eq:E_H}
\end{align}
\GSAedit{This equation was found after assuming isotropic electron
pressure and the equation of state introduced in Sec.
\ref{Sec:Model}}. \GSAedit{However, the contribution of the last
term in Eq. \eqref{Eq:E_H} vanishes once it is inserted in Eq.
\eqref{Eq:vector_V}}.

Finally, note that Eq. \eqref{Eq:FLR_work:3} involves all the components of $\bar{\bm{\Pi}}$. The FLR pressure tensor is symmetric, hence it has six different components. The first row/column is given by Eq. \eqref{e:Pi_ex}. The components $\Pi_{yy}$, $\Pi_{yz}$ and $\Pi_{zz}$ need to be derived from Eqs. \eqref{e:Pi11}-\eqref{e:Pi31}. Similarly to Eq. \eqref{e:Pi_ex}, they can be expressed as
\begin{equation}
    \Pi_{yy} \mathbf{e}_x + \Pi_{yz} \mathbf{e}_y + \Pi_{zz} \mathbf{e}_z =  \frac{1}{M_A \cos \theta} \frac{1}{\hat{b}} \bar{\mathbf{N}}
    \cdot \frac{\partial \mathbf{u}}{\partial \hat{x}}
\end{equation}
with the components of $\bar{\mathbf{N}}$ given by
\begin{align}
& N_{11} = -\hat{b}_x \hat{b}_y \hat{b}_z \left( P_\perp - 4 a_p P_\parallel \right) \\[3mm]
& N_{12} = \frac{1}{2} \hat{b}_z P_\perp \left(1 + 3 \hat{b}_y^2\right) \\[3mm]
& N_{13} = \frac{1}{2} \hat{b}_y \left[ P_\perp \left( 4 + \hat{b}_x^2 - 4 \hat{b}_y^2 - \hat{b}_z^2 \right) - 8 a_p P_\parallel \hat{b}_x^2 \right]\\[3mm]
& N_{21} = \frac{1}{4} \hat{b}_x \left[ 2P_{\perp} \left( \hat{b}_y^2 - \hat{b}_z^2 \right) + 8 a_p P_\parallel \left( \hat{b}_z^2 - \hat{b}_y^2 \right) \right] \\[3mm]
& \nonumber N_{22} = -\frac{1}{4} \hat{b}_y \left[ P_{\perp} \left( 5 + \hat{b}_x^2 - \hat{b}_y^2 - 7 \hat{b}_z^2 \right) \right. - \\ &\qquad \left. - 8 a_p P_\parallel \hat{b}_x^2 \right] \\[3mm]
& N_{23} = \frac{1}{4} \hat{b}_z \left[ P_{\perp} \left( 5 + \hat{b}_x^2 - 7 \hat{b}_y^2 - \hat{b}_z^2 \right) \right. - \nonumber \\ &\qquad \left. - 8 a_p P_\parallel \hat{b}_x^2 \right] \\[3mm]
& N_{31} = \hat{b}_x \hat{b}_y \hat{b}_z \left( P_\perp - 4 a_p P_\parallel \right) \\[3mm]
& N_{32} = -\frac{1}{2} \hat{b}_z \left[ P_\perp \left( 4 + \hat{b}_x^2 - \hat{b}_y^2 - 4 \hat{b}_z^2 \right) \right. - \nonumber\\ &\qquad\left. - 8 a_p P_\parallel \hat{b}_x^2 \right]\\[3mm]
& N_{33} = -\frac{1}{2} \hat{b}_y P_\perp \left(1 + 3 \hat{b}_z^2\right)
\end{align}

 % \textcolor{red}{Estas ecuaciones son las que Thierry nos mandó en su día. He cambiado la notación para que concuerde con la que usamos nosotros (la del paper de Mjolhus). Hay algunos cambios en la posición de la densidad por erratas que ya discutí con Thierry —en el código de Niza las ecuaciones están bien. Por último, comprobé también en su día que son perfectamente equivalentes a las de Mjolhus.}

% \bibliography{biblio}

%

\end{document}